\documentclass[sigconf,edbt]{acmart-edbt2020}
\pdfoutput=1
\usepackage{balance} 
\usepackage{graphicx}
\usepackage{slashbox}
\usepackage{pict2e}
\usepackage[utf8]{inputenc}
\usepackage{url}
\usepackage{amsmath}
\usepackage[T1]{fontenc}
\usepackage{multirow}
\usepackage{adjustbox}

\usepackage{amssymb}
\usepackage{pifont}
\newcommand{\cmark}{\ding{51}}%
\usepackage{mathtools,etoolbox}
\DeclarePairedDelimiterX{\abs}[1]{\lvert}{\rvert}{\ifblank{#1}{{}\cdot{}}{#1}}

\newtheorem{defn}{Definition}

\newcommand{\reffig}[1]{Fig.~\ref{#1}}
\newcommand{\reftab}[1]{Table~\ref{#1}}
\newcommand{\refsec}[1]{Section~\ref{#1}}

\def\BibTeX{{\rm B\kern-.05em{\sc i\kern-.025em b}\kern-.08em
    T\kern-.1667em\lower.7ex\hbox{E}\kern-.125emX}}

\setcopyright{rightsretained}

\acmDOI{}

\acmISBN{XXX-X-XXXXX-XXX-X}

\acmConference[ 2020]{}{2020}{} 
\acmYear{2020}

\begin{document}
\title{The Effects of Different JSON Representations  on \\ Querying Knowledge Graphs}
  

\author{Masoud Salehpour}
\affiliation{\institution{University of Sydney}}
\author{Joseph G. Davis}
\affiliation{\institution{University of Sydney}}

\begin{abstract}
Knowledge Graphs (KGs) have emerged as the de-facto standard for modeling and querying datasets with a graph-like structure in the Semantic Web domain. Our focus is on the performance challenges associated with querying KGs. We developed three informationally equivalent JSON-based representations for KGs, namely, Subject-based Name/Value (JSON-SNV), Documents of Triples (JSON-DT), and Chain-based Name/Value (JSON-CNV). We analyzed the effects of these representations on query performance by storing them on two prominent document-based Data Management Systems (DMSs), namely, MongoDB and Couchbase and executing a set of benchmark queries over them. We also compared the execution times with row-store Virtuoso, column-store Virtuoso, and \mbox{Blazegraph} as three major DMSs with different architectures (aka, RDF-stores). Our results indicate that the representation type has a significant performance impact on query execution. For instance, the JSON-SNV outperforms others by nearly one order of magnitude to execute subject-subject join queries. This and the other results presented in this paper can assist in more accurate benchmarking of the emerging DMSs. 
\end{abstract}

\maketitle

\section{Introduction}
\label{sec::introduction}
Knowledge Graphs (KGs) are logical and semantically rich data models able to represent real-world entities and their interconnections~\cite{SemanticNet,Peng,GenKG,QA,Exfact,Erham1}. Due to their flexibility and expressivity, KGs can be used in a wide range of domains including computer science, medicine, and biology, among others. Several KGs such as Wikidata, YAGO, and Bio2RDF, to name a few, are openly available with new content being added continually~\cite{edbtkg2019,edbtyago2019}. As well, many private organizations such as Amazon, Google, Facebook, and Alibaba have created KGs for different purposes ranging from \textit{semantic search} and \textit{recommendations} to \textit{reasoning}.

Based on the W3C recommendations, KGs are usually represented using Resource Description Framework (\textit{RDF}\footnote{\url{https://www.w3.org/RDF/}}). RDF is a directed, labeled graph-like structure for representing the content of a KG using a set of triples of the form <subject predicate object>. RDF represents subjects and objects of triples as vertices of a graph that are connected by predicates as labeled edges~\cite{IBMapple,QR,edbtcore2019,lodbook}. RDF offers a simple representation for KGs. This simplicity can help provide an intuitive conceptualization of the real-world entities and their inter-relationships in many applications. However, RDF's \textit{flexibility}, absence of an \textit{explicit schema}, and the \textit{heterogeneity} of KG content pose a challenge to Data Management Systems (DMSs) for \textit{querying} KGs efficiently since DMSs typically cannot make any \mbox{a priori} assumption about the structure of the KG content~\cite{saleem,IBMapple}.

DMS designers have employed a variety of \textit{design choices and architectures} to tackle these challenges for querying KGs. For example, a variety of exhaustive indexing strategies~\cite{Thomasindex}, compression techniques, and dictionary encoding (i.e., to keep space requirements reasonable for excessive indexing) have been implemented by major native RDF-stores such as multiple bitmap indexes of Virtuoso or dictionary-based lexical values encoding of Blazegraph. A number of research prototypes have also been presented. For instance,~\cite{tamer2019} proposed a workload-adaptive and self-tuning RDF-store using physical clustering of the underlying data and~\cite{RDF3x} proposed an architecture, namely, `RISC-style' to leverage multiple query processing algorithms and optimization. However, the problem of storing and querying KGs efficiently continues to challenge DMS designers~\cite{saleem,edbtkg2019,edbtcore2019,bonifatibook}.

Through this paper we hope to initiate a discussion on the efficacy of diverse JSON-based representations for KGs and employing document-stores for executing queries. Document-stores are typically the dominant option in the context of web applications.~\cite{Jignesh} presented a research prototype to enable document-stores to query JSON data in a relational database system through a mapping layer. However, to the best of our knowledge, the efficiency of employing JSON-based representations and document-stores for KGs data management purposes (i.e., storage and querying) has not received much research attention. In this paper, we explore the performance effects of query execution using different JSON representations at different scales.

\textit{Syntactically}, any given KG as a set of RDF triples can be represented as \textit{JSON documents}~\cite{AlexRDF,Arash}. In this paper, we synthesized three distinct JSON representations for KGs, namely, Subject-based Name/Value (JSON-SNV), Documents of Triples (JSON-DT), and Chain-based Name/Value (JSON-CNV). We examined the effects of these representations on query performance by loading them on two prominent document-based DMSs, namely, MongoDB and Couchbase and executing a set of benchmark queries over them. In order to provide a comparative perspective, we also analyzed the execution times with row-store Virtuoso, column-store Virtuoso, and \mbox{Blazegraph} as three major DMSs with different architectures. We ran our experiment using three well-known and publicly available benchmarks, namely, \textit{BSBM}, \textit{FishMark}, and \textit{WatDiv} at different \textit{scales}. The benchmark queries were executed over each of the JSON representations separately and query execution times computed to analyze the effects of the different representations on KG query performance. Our results indicate that the representation type has a significant performance impact on query execution.


Our contributions include:
\begin{itemize}
    \item Proposing three distinct JSON-based representations, namely, Subject-based Name/Value (JSON-SNV), Documents of Triples (JSON-DT), and Chain-based Name/Value (JSON-CNV) to store KGs using document-stores and execute queries over them

    \item Comparative performance analysis and experimental evaluation of the proposed representations with row, column, and graph DMSs in supporting the different KG query types
    
    \item Providing explanations for the apparent advantages of different JSON representations depending on the types of queries. These appear to be data locality and lower memory usage (i.e., because of lesser memory allocation to intermediate results). We also speculate that such locality leads to more optimal CPU cache (i.e., L2 cache) utilization.

    \item Communicating clear scientific and practical guidelines to researchers and practitioners through summarizing the lessons learned from our journey and discussing some of the limitations.
\end{itemize}

\begin{figure}[t]
\centering\includegraphics[width=0.5\textwidth]{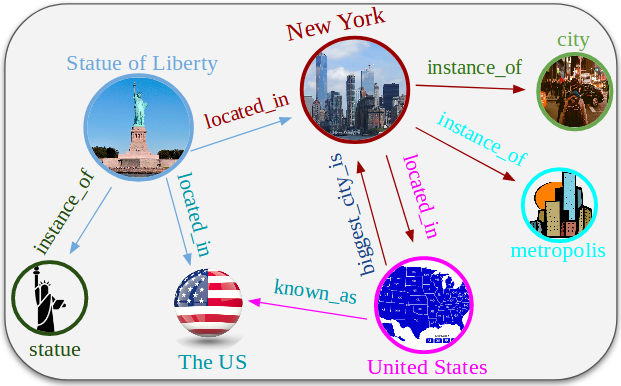}
\caption{A simple Knowledge Graph describing \mbox{Statue of Liberty}}
\label{fig::kg}
\end{figure}

The remainder of this paper is organized as follows. In~\refsec{sec::data_modeling}, we discuss the details of the different JSON representations for KGs.~\refsec{sec::Experimental-Setting} presents our experimental setup including the KG benchmark characteristics, computational environment, DMSs configuration, indexing, data loading process, and query implementation details. In~\refsec{sec::Evaluation}, results of the query processing and related analyses are presented. We summarize the lessons learned from our research and discuss some of the limitations in \refsec{sec::Discussion}.~\refsec{sec::related_work} highlights related work. We present our conclusions and future work in~\refsec{sec::conclusion}.

\section{Knowledge Graph Representations}
\label{sec::data_modeling}
 In this section, we define the key concepts, e.g., a triple, a KG, a JSON document, etc. We also proceed to discuss the three distinct RDF/JSON representations that we have synthesized.

\begin{figure*}[t]
\centering\includegraphics[width=0.8\textwidth]{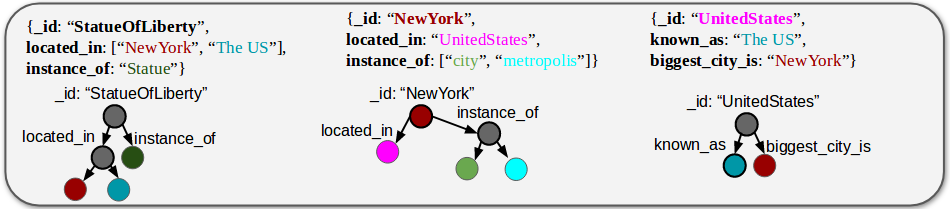}
\caption{JSON documents describing the Subject-based Name/Value (JSON-SNV) representation for the \mbox{Statue Of Liberty} KG (the \reffig{fig::kg})}
\label{fig::jsonsnv}
\end{figure*}

\subsection{Preliminaries}
Assume that $E$ is a set of real-world entities (e.g., ``New York'') and $L$ is a set of literals (e.g., numerical values, dates, etc.). Similar to~\cite{edbtkg2019,xmlbook,CharacteristicsRDF,BinaryRDF,Formaldef2,Formaldef1,Inferray}, we define the key concepts as follows:

\begin{defn}[A Triple Pattern]
(s, p, o) $\in$ E $\times$ (E $\cup$ L) where (s, p, o) is called a triple, in which $s$ is the subject, $p$ the predicate, and $o$ the object.
\end{defn}

\begin{defn}[Knowledge Graph]
A Knowledge Graph KG can be defined as a set of triples where each (s, p, o) is represented as a direct
edge-labelled graph s$\xrightarrow[\text{}]{\text{p}}$o, where:
    \begin{itemize}
    \renewcommand\labelitemi{--}
        \item s , o are vertices of a KG
        \item p is the label for the directed edge of a KG
        \item KG edges $\subseteq (s \times p \times o) $
    \end{itemize}
\end{defn}

\begin{defn}[A Basic Graph Pattern]
A Basic Graph Pattern (BGP) is a subset of Triple Patterns of a Knowledge Graph.
\end{defn}

\reffig{fig::kg} depicts an example of a KG to describe the fact that the \mbox{``Statue of Liberty''} is a heritage site located in \mbox{``New York''}. In this figure, ``Statue of Liberty'' is a subject, ``located\_in'' is a predicate, and ``New York'' is an object. The content of this KG can be represented by the following triples\footnote{We use human-readable names in our examples in this paper. However, Universal Resource Identifiers (URIs), which look like URLs and often include unique sequences of characters and numbers, are generally used to represent triples.}:

\begin{verbatim}
    StatueOfLiberty  located_in      `NewYork'
    StatueOfLiberty  located_in      `The US'
    StatueOfLiberty  instance_of     `Statue'
    NewYork          instance_of     `city'
    NewYork          located_in      `UnitedStates'
    NewYork          instance_of     `metropolis'    
    UnitedStates     known_as        `The US'
    UnitedStates     biggest_city_is `NewYork'
\end{verbatim}

\bigskip

\noindent We define a JSON document with respect to JSON grammar\footnote{The JSON grammar is defined in full detail at: https://www.json.org/} as follows:

\begin{defn}[JSON Document]
Given a collection of name ($N$) and value pairs ($V$), a JSON document $D$ over $N$ and $V$ is a structure as follows:
\begin{center}
    $D = (n_i, v_i)$
\end{center} 
where names ($n_i$) are identifiers and each value ($v_i$) can be:
    \begin{itemize}
    \renewcommand\labelitemi{--}
        \item An atomic value (e.g., a number, text, etc.)
        \item Another JSON object
        \item An ordered list (i.e. an array) of values
    \end{itemize}
Each JSON document can be uniquely identified by a distinct name/value pair usually called the ``id'' pair (i.e., the root of the tree).
\end{defn}

JSON is a flexible data serialization format~\cite{JSONTree}. As the W3C recommended,\footnote{\url{https://www.w3.org/2018/jsonld-cg-reports/json-ld/}} KGs can be \textit{syntactically} represented using the JSON data format~\cite{AlexRDF,Arash}. In the following, we proceed to discuss the three distinct and informationally equivalent JSON representations that we developed for storing KGs and querying over them.

\subsection{Subject-based Name/Value (JSON-SNV)}
\label{sec::snv}
In the Subject-based Name/Value (JSON-SNV) representation, a JSON document $D = (N, V)$ consists of a name set $N$ and a value set $V$. A name $n \in N$ appears in a name/value pair somewhere in a JSON document. A value $v \in V$ is from a name/value pair such that, $n: v$ means $v$ is the value for the name $n$. In this representation, the ``id'' pair is organized based on the subject of a set of triples. $N$ is the set of predicates that are associated with the subject, and $V$ is the set of associated objects. \reffig{fig::jsonsnv} shows the structure of the JSON-SNV for the KG depicted in \reffig{fig::kg}. This representation is subject-based in the sense that the ``New York'' as a value is associated with the ``id'' name. The ``New York'' is a subject and has three predicates, therefore, the JSON representation has four names (i.e., three predicates + the ``id'' name/value pair). Objects appear as values associated with predicates directly or inside an array. This representation is similar to the JSON-LD\footnote{https://json-ld.org/} and Talis' RDF/JSON~\cite{AlexRDF} which are used in Google Knowledge Graph.\footnote{https://developers.google.com/knowledge-graph/} The JSON-SNV can be defined formally as follows:

\begin{defn}[Subject-based Name/Value]
Assume that $G$ is a KG and $S_G$, $P_G$, $O_G$ are sets of subjects, predicates, objects in the $G$, respectively. Let $(s,p,o)$ be a triple of the $G$ such that $s \in S_G, p \in P_G$, and $o \in O_G$. Assume that $U_G$ is the set of $(predicates, objects)$. A JSON-SNV document is then a representation for a KG, whose name/value pairs meet the following criteria:
\begin{itemize}
    \renewcommand\labelitemi{--}
\item $id \in S_G$ (i.e., ``id'' of a JSON documents should be associated with a $unique$ subject of a triple)
        \item $U_G = \{ (p,o) | \exists u:(s,p,o) \in G \}\ $ where each $p$ is a name of a JSON document and each $o$ is the associated value.
        \item $O_G$ appear as values associated with predicates (i.e., $P_G$) directly or inside an array.
    \end{itemize}
\end{defn}

\begin{figure*}[t]
\centering\includegraphics[width=0.8\textwidth]{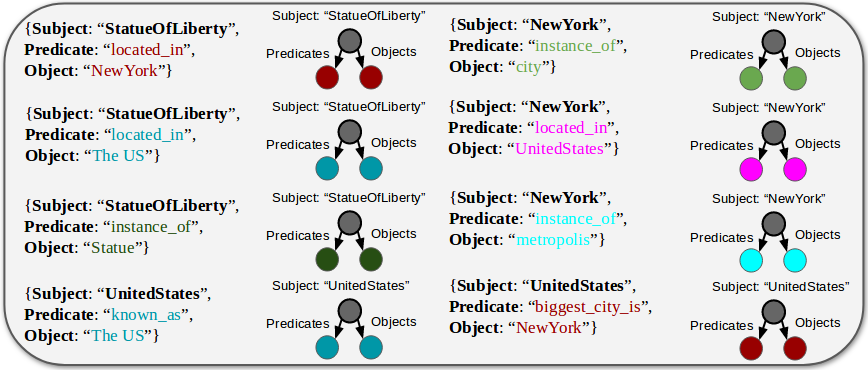}
\caption{JSON documents describing the Documents of Triples (JSON-DT) representation for the \mbox{StatueOfLiberty} KG (the \reffig{fig::kg})}
\label{fig::triplesub}
\end{figure*}

\subsection{Documents of Triples (JSON-DT)}
In the Documents of Triples (JSON-DT) representation, a JSON document $D = (s,p,o)$ consists of three names and the associated values. ``Subject'', ``Predicate'' and ``Object'' are three names which appear in all JSON documents in the JSON-DT representation. In each JSON document, these names are associated with $s$ the subject, $p$ the predicate, and $o$ the object of a triple, respectively. \reffig{fig::triplesub} shows the structure of a JSON-DT representation for the KG depicted in \reffig{fig::kg}. We can see that ``Statue Of Liberty'' as a value is associated with the ``Subject'' name. ``Predicate'' and ``Object'' appear in all JSON documents and the associated values (i.e., $p$ and $o$) are paired directly with them (no array is used in the JSON-DT representation). This representation is similar to the natural representation of triples as defined by W3C.\footnote{https://www.w3.org/TR/n-triples/} This representation can be defined formally as follows:

\begin{defn}[Documents of Triples]
Let $G$ be a KG and $S_G$, $P_G$, $O_G$ be sets of subjects, predicates, objects in the $G$, respectively. Assume that $(s,p,o)$ is a triple of the $G$ such that $s \in S_G$, $p \in P_G$, and $o \in O_G$. Then, a JSON-DT document is a representation for a KG, whose name/value pairs meet the following criteria:
\begin{itemize}
    \renewcommand\labelitemi{--}
\item $ Subject = \{ (S_G) | \exists s:(s,p,o) \in G \}$ 
        \item $ Predicate = \{ (P_G) | \exists p:(s,p,o) \in G \}$
        \item $   Objects = \{ (O_G) | \exists o:(s,p,o) \in G \}$
\end{itemize}
\end{defn}

\subsection{Chain-based Name/Value (JSON-CNV)}
\label{subsec::CNV}
\begin{figure*}[t]
\centering\includegraphics[width=0.8\textwidth]{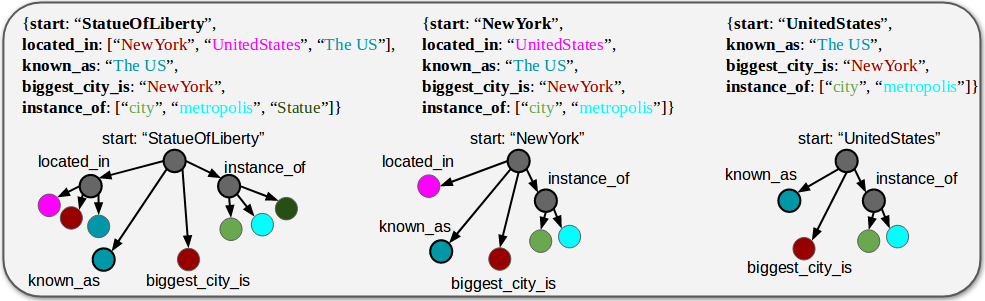}
\caption{JSON documents describing the Chain-based Name/Value (JSON-CNV) representation for the \mbox{StatueOfLiberty} KG (the \reffig{fig::kg})}
\label{fig::triplecnn}
\end{figure*}

In KGs, the object of a triple can itself be the subject of another triple. The information of such sequences of connected subjects and objects is appeared in JSON documents in Chain-based Name/Value (JSON-CNV) representation. In other words, graph paths are first-class citizens (along with the subjects and objects) in this model. \reffig{fig::triplecnn} shows the structure of this representation for the KG depicted in \reffig{fig::kg}. In this example, three JSON documents are represented the \mbox{``Statue Of Liberty''} KG (the \reffig{fig::kg}). Obviously, KGs can have multiple paths which can represent redundant or overlapping edge sequences. For instance, such an overlap can be seen in~\reffig{fig::triplecnn} where ``biggest\_city\_is'' is appeared in all JSON documents. This is inspired by the concept of objectified paths for graphs~\cite{bonifatibook}. This representation can be defined formally as follows:

\begin{defn}[Chain-based Name/Value]
Let $G$ be a KG and $S_G$, $P_G$, $O_G$ be sets of subjects, predicates, objects in the $G$, respectively. Then, a JSON-CNV document can be represented with the following structure:
\begin{center}
    D = ($S_G$, $P_G$, $C_G$, $\theta$, $\kappa$)
\end{center}
where
\begin{itemize}
    \renewcommand\labelitemi{--}

\item  $C_G \subset O_G$ is a finite set of connected objects, also called a path between a sequence of objects.
        \item $\theta$:   $C_G \rightarrow \bigcup\limits_{n \geq 0}^{} P_{G}^n$ is a total function assigning to every path a sequence of predicates
        \item $\kappa$: $S_G \cup P_G \cup C_G  \rightarrow \Omega (\Lambda)$ is a function inserting objects of each chain and their associated values to a JSON document
        \end{itemize}
        such that
        
        \begin{itemize}
    \renewcommand\labelitemi{--}
        \item sets $S_G$ and $O_G$ may have some overlaps
        \item ($S_G \cup O_G$) $ \cap P_G = \emptyset $  
        \end{itemize}
\end{defn}

\subsection{Informational Equivalence}
\begin{figure}[t]
\centering\includegraphics[width=0.5\textwidth]{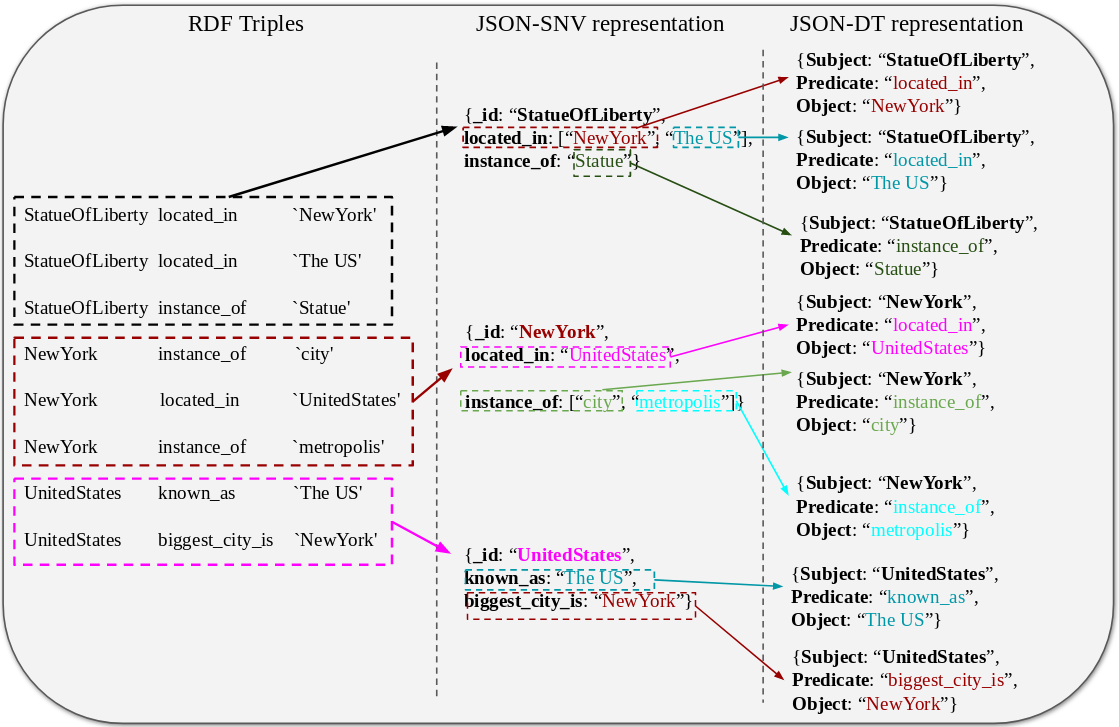}
\caption{A matching between instances of JSON-SNV and JSON-DT representation}
\label{fig::snv2dt}
\end{figure}

We draw on the literature~\cite{xmlbook,fagin,Fuxman} on schema mapping of relational/XML data models to show the informational equivalence among the three JSON representations. Such mapping describes connections between JSON representations to determine whether they reflect the same source data. We define the information equivalence of JSON documents as follows:

\begin{defn}[Information Equivalence of JSON Representations]
Two different JSON representations of the same KG will be called informationally equivalent when it is possible to map the schema of one of them to another. The schema mapping process consists of:
   \begin{itemize}
    \renewcommand\labelitemi{--}
        \item A source schema
        \item A target schema
        \item A set of source-to-target schema mapper of the form:
        \end{itemize}
        \begin{center}
            $\pi(\overline{n},\overline{v}) \rightarrow \exists \pi^\prime(\overline{n},\overline{v})  $
        \end{center}
        where $\pi$ and $\pi^\prime$ are two different JSON documents that usually contain the same pair of name/value using different structures.
        
\end{defn}
 \reffig{fig::snv2dt} illustrates an example for showing information equivalence between JSON-SNV and JSON-DT. This figure has three representations: The left-most depicts the RDF triples related to the \mbox{``Statue Of Liberty''} KG, the JSON-SNV representation is shown in the middle, and the right-most is the JSON-DT. The dotted lines show how the same information is structured using these representations. Similarly, \reffig{fig::snv2dt} shows another example in which instances of JSON-SNV and JSON-CNV are matched. There may be some redundant or overlapping name/value pairs in the representations. However, the different JSON representations include the same information with or without redundancy.

\subsection{Knowledge Graph Query Types}
\label{subsec::kgqueries}
\noindent KG queries contain a set of BGPs (please see Definition 3) which are like KG triples. In each BGP, the subject, predicate, and object can be a variable. An example of a KG query\footnote{We assume that the reader is familiar with the basic concepts of querying KG, e.g., the SELECT clauses.} with a BGP is given below:
\begin{verbatim}
    SELECT ?Ins
    WHERE {
    StatueOfLiberty instance_of ?Ins .
    }
\end{verbatim}

This KG query asks for the subject ``Statue of Liberty's'' type (from the KG in \reffig{fig::kg}). It has one BGP where its object is a variable to return the associated value as the result (i.e., ``statue''). KG queries can generally be defined formally as follows:

\begin{defn}[Knowledge Graph Query]
Assume that G is a KG, and $\mu$ is the resultset of matching a BGP against G such that $\mu(BGP)$ is a subgraph of G containing triples that are matched the query variables in the BGP:

\begin{itemize}
    \item $\mu(BGP)$ = (all triples that are matched the query variables in BGP, such that $\mu(BGP)$ contains a subgraph of G including all matched triples).
    \item $\abs{\mu(BGP)}$ = (cardinality of $\mu(BGP)$, i.e., number of matched triples in the subgraph G).
\end{itemize}
\end{defn}


\textbf{Join Queries.} As explained above, each BGP of a KG query returns a subgraph. This resultant subgraph can be further \textit{joined} with the results of other BGPs in the query to return the final resultset. In practice, there are three major types of join queries: (i) Subject-subject joins, (ii) subject-object joins, and (iii) combination of subject-subject and subject-object joins~\cite{Survey2018,saleem,sakrbook}. These types of join queries are explained in the following.

\textbf{Subject-subject Joins.} A subject-subject join is performed by a DMS when a KG query has at least two BGPs such that the predicate and object of each BGP is a given value, but the subjects of both BGPs are replaced by the same variable. For example, the following query looks for all subjects of the KG in~\reffig{fig::kg} that are: located in ``The US'' and instance of ``statue'' (the result will be ``Statue of Liberty'').

\begin{verbatim}
    SELECT ?x
    WHERE {
    ?x  located_in   "The US" .
    ?x  instance_of  "statue" .
    }
\end{verbatim}
Subject-subject joins can be defined formally as follows:

\begin{defn}[A Subject-subject Join]
Assume that $\lambda_1$ and $\lambda_2$ are resultsets of $\mu(BGP_1)$ and $\mu(BGP_2)$, respectively (i.e., $\lambda_1 = \mu(BGP_1)$ and $\lambda_2 = \mu(BGP_2)$). We define a subject-subject join as follows:
\begin{itemize}
    \item Join($\lambda_1, \lambda_2$) = merge($\mu(BGP_1)$, $\mu(BGP_2)$) $\bigm\lvert$ $\mu(BGP_1)$ in $\lambda_1$ and  $\mu(BGP_2)$ in $\lambda_2$, and $\mu(BGP_1)$ and $\mu(BGP_2)$ are compatible)
    \item merge($\lambda_1$, $\lambda_2$) only consists of those triples such that the subject of all the triple patterns involved in the query
\end{itemize}
\end{defn}

\textbf{Subject-object Joins.} A subject-object join is performed by a DMS when a KG query has at least two BGPs such that the subject of one of the BGPs and the object of the other BGP are replaced by the same variable. For example, the following query looks for all subjects that are located within American cities (the result will be ``Statue of Liberty'').

\begin{verbatim}
    SELECT ?y
    WHERE {
    ?x  located_in  "United States"  .
    ?y  located_in  ?x               .
    }
\end{verbatim}
Subject-object joins can be defined formally as follows:

\begin{defn}[A Subject-object Join]
Assume that $\sigma_1$ and $\sigma_2$ are resultsets of $\mu(BGP_1)$ and $\mu(BGP_2)$, respectively (i.e., $\sigma_1 = \mu(BGP_1)$ and $\sigma_2 = \mu(BGP_2)$). We define a subject-object join as follows:
\begin{itemize}
    \item Join($\sigma_1, \sigma_2$) = { merge($\mu(BGP_1)$, $\mu(BGP_2)$) | $\mu(BGP_1)$ in $\sigma_1$ and  $\mu(BGP_2)$ in $\sigma_2$, and $\mu(BGP_1)$ and $\mu(BGP_2)$ are compatible}
    \item merge($\sigma_1$, $\sigma_2$) only consists of those triples patterns which  are consecutively connected like a chain within a KG
   
\end{itemize}
\end{defn}

\textbf{Combinations of Subject-subject and Subject-object Joins.} This query type consists of \textit{combinations} of subject-subject and subject-object joins. In other words, these queries consist of several subject-subject sub-queries connected via some subject-object join queries.

\section{Experimental Setup}
\label{sec::Experimental-Setting}
\noindent In this section, we describe the benchmark KGs. As well, our \textit{computational environment} and the DMS \textit{configurations} are described in detail.

\subsection{Knowledge Graph Benchmarks}
We used three well-known benchmarks in this research. These are publicly available KG datasets with a collection of benchmark queries. These benchmarks are also recognized as major KG querying benchmarks by previous studies such as~\cite{saleem,sakrbook,ISWC2013}. These benchmarks are as follows:

\textbf{Berlin SPARQL Benchmark\footnote{\url{http://wifo5-03.informatik.uni-mannheim.de/bizer/berlinsparqlbenchmark/}} (BSBM)~\cite{bsbm}.} The BSBM benchmark is built around an e-commerce use-case in which a set of products is offered by different vendors and consumers have posted reviews of these products on various review sites. The BSBM benchmark queries emulate the search and navigation pattern of typical users looking for products with similar properties. The benchmark follows specific rules that allow us to scale benchmark KGs to arbitrary sizes using the number of products as the scale factor. We used this benchmark KG in three different sizes, namely, 10 \textbf{M}illion, 100M, and 1000M triples.

\textbf{Waterloo SPARQL Diversity Test Suite\footnote{\url{https://dsg.uwaterloo.ca/watdiv/}} (WatDiv)~\cite{watdiv}.} The WatDiv benchmark is designed to generate heterogeneous RDF datasets in which subjects of the same type (e.g., website or retailer) usually do not have the same predicates and objects. In other words, the WatDiv benchmark is designed to measure how a DMS for RDF datasets performs against queries with varying structural and selectivity characteristics. We used this benchmark KG in three different sizes, namely, 10M, 100M, and 1000M triples.

\begin{figure}[t]
\centering\includegraphics[width=0.5\textwidth]{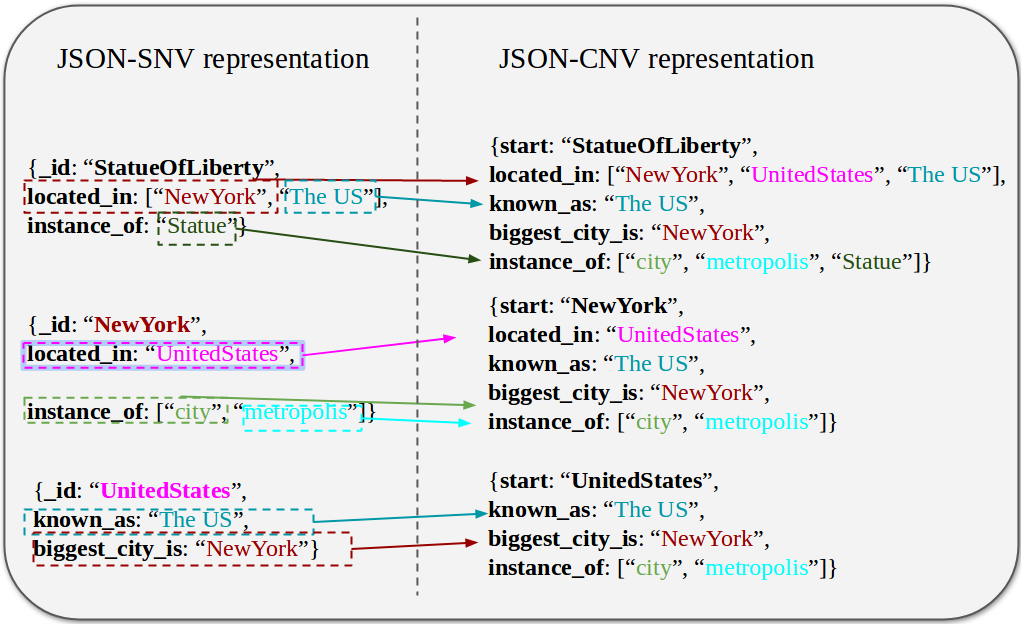}
\caption{A matching between instances of JSON-SNV and JSON-CNV representation}
\label{fig::snv2cnv}
\end{figure}

\begin{table*}[t]
\centering
\begin{tabular}{ccccccc}
\hline
 \backslashbox{Benchmark}{Statistics} & Scale (nominal) & Subjects (\#) & Predicates (\#)& Objects (\#) & Triples (\#)\\
 \hline\hline

\multicolumn{1}{ c  }{\multirow{3}{*}{BSBM} } &
\multicolumn{1}{ c }{10M} & 934,324 & 40 & 1,919,901 & 10,190,687    \\ 
\multicolumn{1}{ c  }{}                        &
\multicolumn{1}{ c }{100M} & 9,197,305 & 40 & 15,207,734 & 100,652,457    \\ 
\multicolumn{1}{ c  }{}                        &
\multicolumn{1}{ c }{1000M} & 91,647,129 & 40 & 140,996,171 & 1,004,406,629     \\ \cline{1-6}
\hline
\hline
\multicolumn{1}{ c  }{\multirow{3}{*}{WatDiv} } &
\multicolumn{1}{ c }{10M} & 521,585 & 86 & 1,005,832 & 10,916,457 \\
\multicolumn{1}{ c  }{}                        &
\multicolumn{1}{ c }{100M} & 5,212,385 & 86 & 9,753,266 & 108,997,714 \\ 
\multicolumn{1}{ c  }{}                        &
\multicolumn{1}{ c }{1000M} & 52,120,385 & 86 & 92,220,397 & 1,092,155,948 \\ \cline{1-6}
\hline
\hline
\multicolumn{1}{ c  }{\multirow{1}{*}{FishMark} } &
\multicolumn{1}{ c }{10M} & 395,491 & 878 & 1,148,159 & 10,002,178    \\ \cline{1-6}

\end{tabular}
\caption{Characteristics of the Benchmark KGs that were used to run the experiments along with detailed statistics depicted from columns 3-6. The first two columns show the name and the nominal size of the KGs followed by the number of unique subjects, predicates, and objects. The last column depicts the total number of triples of each benchmark KG.}
\label{table::kgs}
\end{table*}

\textbf{FishMark~\cite{fishmark}.} FishMark consists of a KG dataset with relevant queries derived from FishBase (i.e., a comprehensive database about the world's fish species and its popular web front end\footnote{\url{http://fishbase.org}} to FishBase). This benchmark is triplified from FishBase into around 10M triples and is publicly available\footnote{\url{https://hobbitdata.informatik.uni-leipzig.de/benchmarks-data/datasets-dumps/}}.

\reftab{table::kgs} shows the statistical information related to the above KG benchmarks. The RDF representations of these benchmarks are available in different formats such as N-Triples and Turtle. We converted these benchmark datasets from RDF/N-Triples syntax to the three JSON collections using a parser designed and developed as part of this project\footnote{The source code is available through \url{https://github.com/m-salehpour/JSON-DM}}. Each JSON collection is equivalent to one of the three representations described in \refsec{sec::data_modeling}. During the conversion, we performed a minor modification of the data by shortening the URIs that were obviously overloaded to correspond to several real-world entities. The semantic network and content (i.e., the total number of triples) of the datasets remain unchanged since we did not change any of the relationships between the triples.

\textbf{Benchmark Queries.} KG benchmarks usually contain four query forms, namely, ``SELECT'',``ASK'',``DESCRIBE'', and ``CONSTRUCT''. These forms are explained in the W3C portal in detail\footnote{\url{https://www.w3.org/TR/sparql11-query/}}. Similar to previous works such as~\cite{saleem,ISWC2013,watdiv,svenbook,Hexastore}, our specific focus is on the performance of ``SELECT'' queries in this paper. We selected 23 representative queries\footnote{All queries are available through \url{https://github.com/m-salehpour/JSON-DM}} across the benchmark KGs. All or some of these queries have also been used in previous studies such as~\cite{saleem,ISWC2013,watdiv}. We ran these benchmark queries against the corresponding datasets using the DMSs. \reftab{table::queries} shows the classification of the 23 queries as SS, SO, Co, and selective (more details in \refsec{subsec::kgqueries}).

\begin{table}[h]
\centering
\begin{tabular}{cccccc}
\hline
 \backslashbox{Benchmark}{Types}& Query & $SS^{*}$ & $SO^{**}$ & $Co^{***}$ & Selective  \\
 \hline\hline

\multicolumn{1}{ c  }{\multirow{8}{*}{BSBM} } &
\multicolumn{1}{ c }{Q1} & \cmark &  &  &   \\ 
\multicolumn{1}{ c  }{}                        &
\multicolumn{1}{ c }{Q2} & \cmark  & & & \cmark   \\ 
\multicolumn{1}{ c  }{}                        &
\multicolumn{1}{ c }{Q4} & \cmark &&&     \\ 
\multicolumn{1}{ c  }{}                        &
\multicolumn{1}{ c }{Q10} & &\cmark&  &    \\ 
\multicolumn{1}{ c  }{}                        &
\multicolumn{1}{ c }{Q11} & & \cmark  & &    \\ 
\multicolumn{1}{ c  }{}                        &
\multicolumn{1}{ c }{Q16} & &\cmark  &&    \\ 
\multicolumn{1}{ c  }{}                        &
\multicolumn{1}{ c }{Q18} & & \cmark  & &   \\ 
\multicolumn{1}{ c  }{}                        &
\multicolumn{1}{ c }{Q20} &\cmark &  &&\cmark     \\ \cline{1-6}
\hline
\hline
\multicolumn{1}{ c  }{\multirow{3}{*}{WatDiv} } &
\multicolumn{1}{ c }{Q1-3} & \cmark  & && \\
\multicolumn{1}{ c  }{}                        &
\multicolumn{1}{ c }{Q4-6} &   & \cmark && \\ 
\multicolumn{1}{ c  }{}                        &
\multicolumn{1}{ c }{Q7-8} &   & & \cmark & \\ \cline{1-6}
\hline
\hline
\multicolumn{1}{ c  }{\multirow{3}{*}{FishMark} } &
\multicolumn{1}{ c }{Q1-2} &   & \cmark && \\
\multicolumn{1}{ c  }{}                        &
\multicolumn{1}{ c }{Q3 \& Q8} &  \cmark &  &&\cmark \\ 
\multicolumn{1}{ c  }{}                        &
\multicolumn{1}{ c }{Q4-7} & \cmark  & &  & \\ \cline{1-6}

\end{tabular}
\caption{Types of the benchmark queries. $SS^{*}$: Subject-subject join, $SO^{**}$: Subject-object join, $Co^{***}$: combination of SS and SO}
\label{table::queries}
\end{table}

\subsection{System Settings}
\label{sec:db}
\textbf{Computational Environment.} Our benchmark system is a Virtual Machine (VM) instance with a 2.3GHz AMD Processor, running Ubuntu Linux (kernel version: 4.4.0-161-generic), with 48GB of main memory, 16 vcores, 512K L2 cache, 5TB instance storage capacity. The VM cache read is roughly 2799.45MB/sec and the buffer read is roughly 35.85MB/sec (i.e., the output of the ``hdparm -Tt'' Linux command). The operating system is set with almost no ``soft/hard'' limit on the file size, CPU time, virtual memory, locked-in-memory size, open files, processes/threads, and memory size using Linux ``ulimit'' settings.

\noindent \textbf{Data Management Systems (DMSs).} We chose five different DMSs as follows: (1) Row-store Virtuoso (Open Source Edition, version 06.01.3127), (2) Column-store Virtuoso (Open Source Edition, Version 07.20.3230--commit 4a668a5), (3) Blazegraph\footnote{Previously known as Bigdata DB.} (Open Source Edition, version 2.1.5--commit 3122706), MongoDB (enterprise edition, version: 4.2.0), and Couchbase (enterprise edition, version: 6.0.2-2413-1). All or some of these DMSs have also been used in previous studies such as~\cite{watdiv,ISWC2013,Medha2,bsbm,biobench,saleem}. We measured the performance of Blazegraph, row-store and column-store Virtuoso as native RDF-stores to serve as the baseline in our experiment. \textit{The goal of this evaluation was not to compare DMSs like document-stores versus RDF-stores but to analyze the effects of the different JSON representations on query performance against the baseline DMSs and to establish a platform for replicable evaluation}.

\noindent \textbf{Configuration of row- and column-store Virtuoso.} We configured both row- and column-store Virtuoso based on the vendor's official performance tuning recommendations\footnote{\url{http://vos.openlinksw.com/owiki/wiki/VOS/VirtRDFPerformanceTuning}}. For example, we configured the Virtuoso process to use the main memory and the storage disk effectively by setting ``NumberOfBuffers'' to ``4,000,000'', ``MaxDirtyBuffers'' to ``3,000,000'', and ``MaxCheckpointRemap'' to ``a quarter'' of the database size as recommended. We also used the latest version of GNU packages that are necessary to build column-store Virtuoso (e.g. GNU gpref 3.0.4, libtool 2.4.6, flex 2.6.0, Bison 3.0.4, and Awk 4.1.3).

\noindent \textbf{Configuration of Blazegraph.} We configured Blazegraph based on the vendor's official performance tuning recommendations\footnote{\url{https://wiki.blazegraph.com/wiki/index.php/PerformanceOptimization}} as well. For example, we ran our experiments in the ``Worm'' standalone persistence store mode. We turned off all inference, truth maintenance, statement identifiers, and the free text index in our experiment since reasoning efficiency was not part of our research focus in this paper.

\noindent \textbf{Configuration of MongoDB.} We used the default settings for MongoDB. We set MongoDB's level of \textit{profiling to ``2''} to log the data for all query-related operations for precise and detailed query execution time extraction.

\noindent \textbf{Configuration of Couchbase.} We used the default settings for Couchbase as well in which there is almost no limit to use the VM resources as a cluster with a single node.

\noindent \textbf{Indexing of Virtuoso.} We did not change the default indexing scheme of \textit{Virtuoso} (row- and column-store). As highlighted in the official website, ``alternate indexing schemes are possible but will not be generally needed\footnote{\url{http://docs.openlinksw.com/virtuoso/rdfperfrdfscheme}}''. More specifically, Virtuoso's (row and column) data modeling is based on a relational table with three columns\footnote{In the case of loading named graphs, it adds another column for the context, called C} for S, P, and O (i.e., S: Subject, P: Predicate, and O: Object) and carrying multiple indexes over that table to provide a number of different access paths. Most recently, column-store Virtuoso added columnar projections to minimize the on-disk footprint associated with RDF data storage. Virtuoso (row and column) creates the following compound indexes by default for the loaded KG: PSO, PO, SP, and OP.

\noindent \textbf{Indexing of Blazegraph.} As recommended in the Blazegraph's official website\footnote{\url{https://wiki.blazegraph.com/wiki/index.php/PerformanceOptimization}}, we did not change its default data modeling or the indexing schema. Blazegraph's data modeling is based on \mbox{B+Trees} to store KGs in the form of ordered data. Blazegraph typically uses the following three indexes based on the stored \mbox{B+Trees} for triples modes: SPO, POS, and OSP. For normal use cases, these indexes are laid out on variable sized pages. These index pages are read from the backing store and load in the main memory on demand (i.e., into the Java heap). However, Blazegraph takes advantage of a variety of data structures to execute queries when stored KG content is loaded in the main memory. For example, the underlying data model (i.e., \mbox{B+Trees}) is retained by a mixture of a ring buffer (hard reference queue), weak references, and hard references on the stack during the use alongside with a native memory cache for buffering writes to reduce write application effects.

\begin{figure*}[t]
\begin{adjustbox}{center}
\centering\begin{tabular}{c}
\includegraphics[width=1\textwidth]{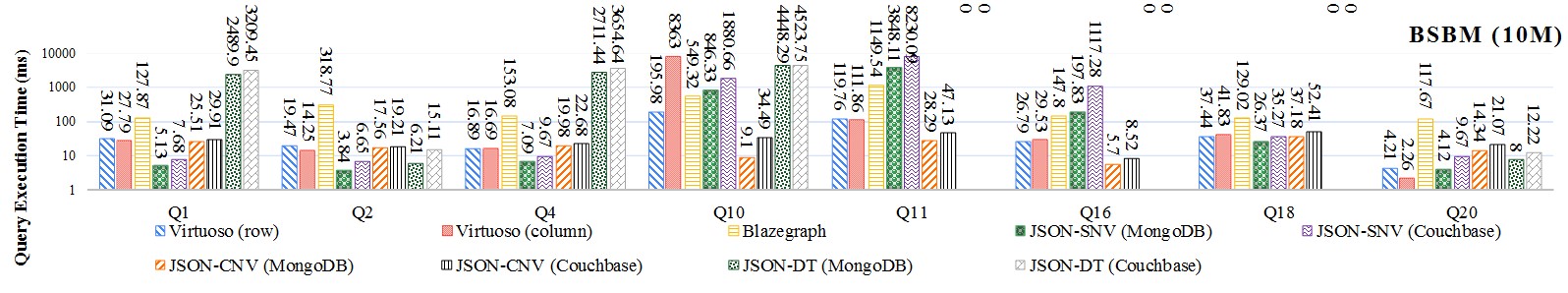}\\
\includegraphics[width=1\textwidth]{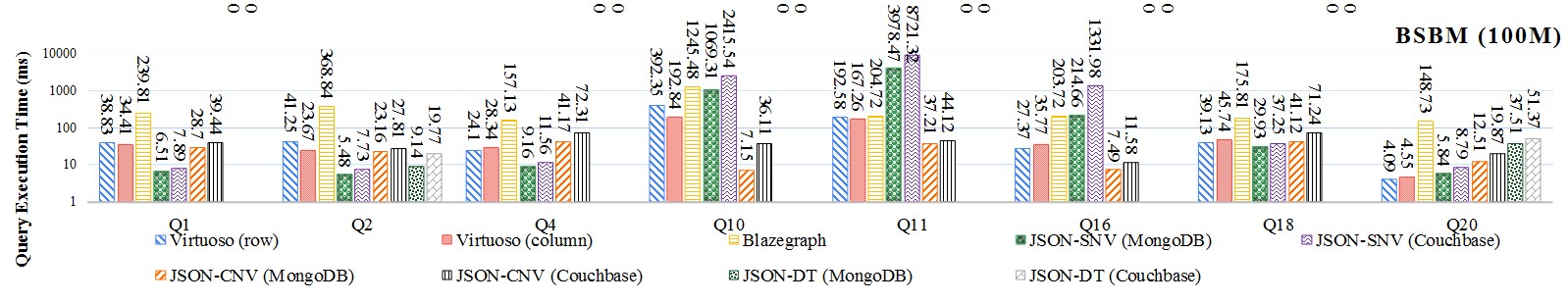}\\
\includegraphics[width=1\textwidth]{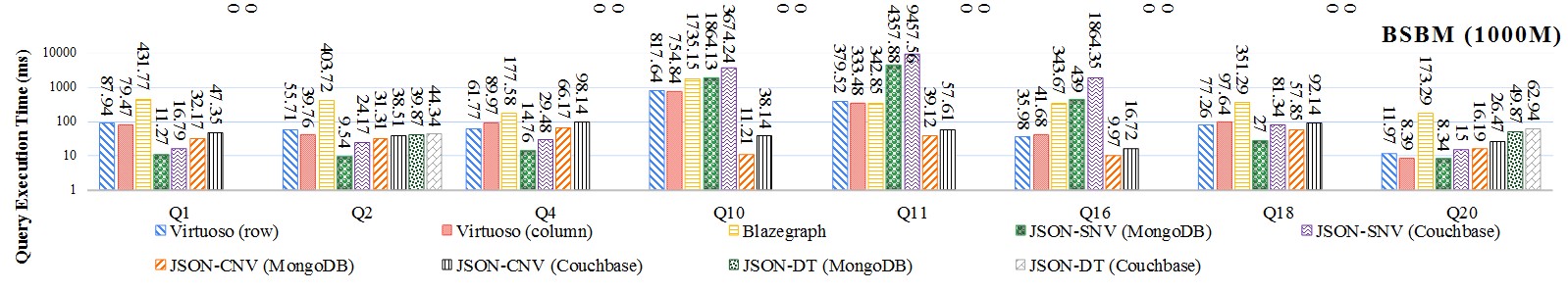}\\
\end{tabular}
\end{adjustbox}
\caption{The BSBM benchmark queries at different scales (i.e., 10M, 100M, and 1000M). $X$ axis shows different benchmark queries. $Y$ axis shows the execution time of each query in milliseconds (log scale). For some queries, the maximum execution time exceeded the time-out threshold which is marked `$0$'.}%
\label{fig::bsbm}%
\end{figure*}

\noindent \textbf{MongoDB and Couchbase Storage Layouts.} We did not change the default storage engines of the document-based DMSs. JSON documents are stored directly by these DMSs\footnote{MongoDB uses the binary equivalent of each JSON document (i.e., BSON) for storage, in which the structure of each document remained unchanged}. Both MongoDB and Couchbase use key/value stores as their internal storage engine, i.e., by default, WiredTiger for MongoDB and ForestDB for Couchbase. These DMSs usually assign an arbitrary (and unique) identifier to each JSON document as a key and consider the document as a value to store them. MongoDB and Couchbase use different variations of \mbox{B-Trees}, to create indexes on the contents of each JSON document.

\noindent \textbf{Indexing of MongoDB.} We created a unique index on those name/value pairs of the JSON representations that were representatives of subjects and predicates.

\noindent \textbf{Indexing of Couchbase.} We created the same indexes as MongoDB for Couchbase. There are no differences between these two document-stores in terms of indexing strategies.

\noindent \textbf{Loading the benchmark KG.} We loaded the RDF/N-Triples format of KG benchmarks into \textit{Virtuoso} (row and column) by using the Virtuoso native bulk loader function (i.e., ``ld\_dir''). To load the KGs into \textit{Blazegraph}, we used Blazegraph's native ``DataLoader'' utility\footnote{\url{https://wiki.blazegraph.com/wiki/index.php/Bulk_Data_Load}}. We loaded KGs into MongoDB using its native tool called ``mongoimport''. In a similar way, we used Couchbase's native tool called ``cbimport'' to load JSON-based representations of KGs.

\noindent \textbf{Shutdown store, clear caches, restart store.} We measured the query execution times in our evaluation. This is end-to-end time computed from the time of query submission to the time when the result is outputted. After the execution of each query, we carefully checked to ensure that the output results are correct and exactly the same across different DMSs. To execute queries over each KG, the DMSs were reinstalled completely and all information related to previous installations was completely removed. For fairness, the query times reported for each DMS are averaged over 5 successive runs (with almost no delay in between) to account for any randomness and noise.

\begin{figure*}[t]
\begin{adjustbox}{center}
\centering\begin{tabular}{c}
\includegraphics[width=1\textwidth]{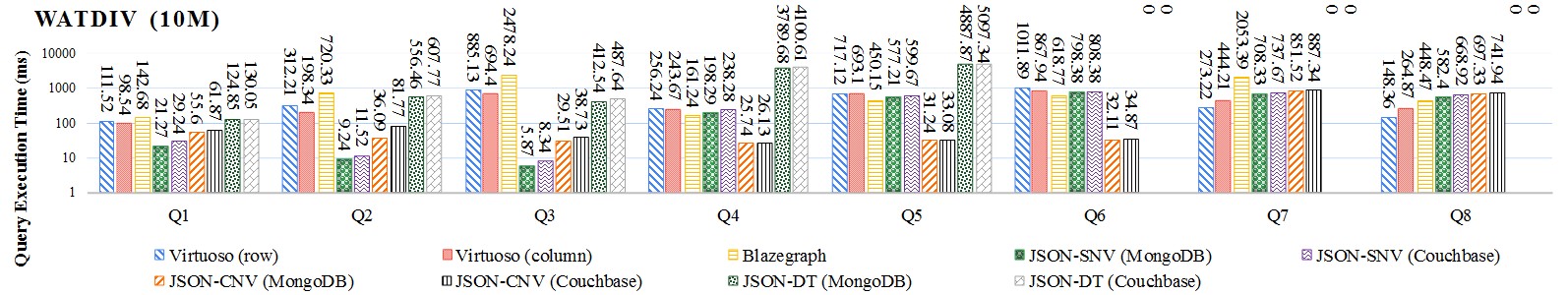}\\
\includegraphics[width=1\textwidth]{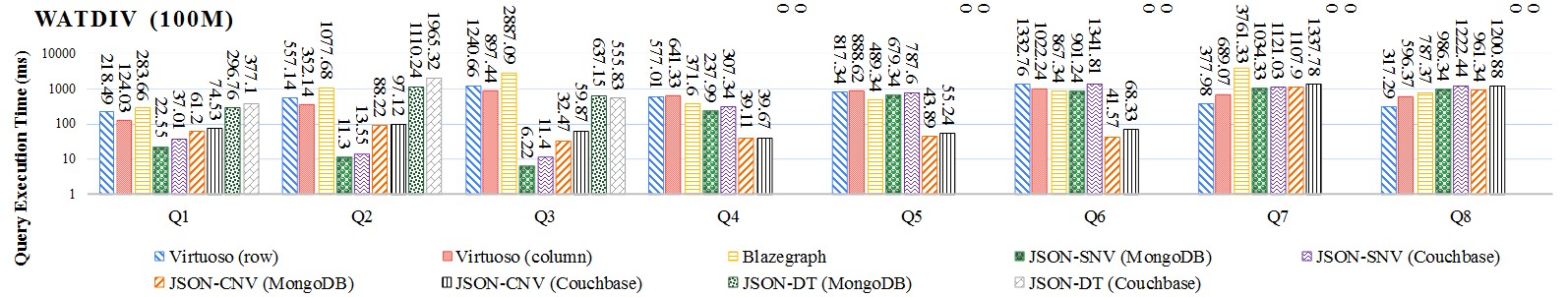}\\
\includegraphics[width=1\textwidth]{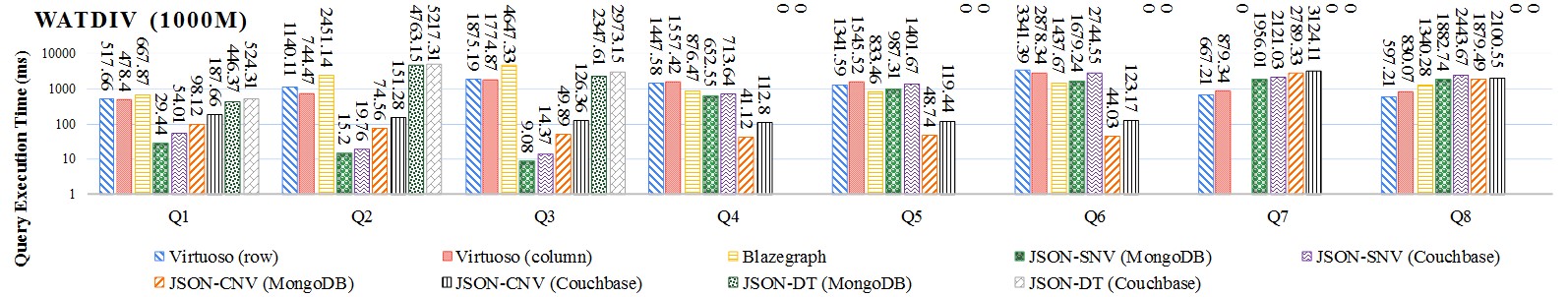}\\
\end{tabular}
\end{adjustbox}
\caption{The WatDiv benchmark queries at different scales (i.e., 10M, 100M, and 1000M). $X$ axis shows different benchmark queries. $Y$ axis shows the execution time of each query in milliseconds (log scale). For some queries, the maximum execution time exceeded the time-out threshold which is marked `$0$'.}%
\label{fig::watdiv}%
\end{figure*}

\section{Evaluation}
\label{sec::Evaluation}
\noindent We evaluated the effects of the JSON-based representations (i.e., JSON-SNV, JSON-DT, and JSON-CNV) on the performance of query execution. Our goal is to understand systematic performance differences, if any, based on the different representations.

\subsection{Results}
\label{sec::results}
We ran the BSBM benchmark queries over the loaded JSON documents that corresponded to KGs of different sizes with 10 \textbf{M}illion, 100M, and 1000M triples. The query execution times over the KGs with different triples are shown in \reffig{fig::bsbm}, in which $X$ axis shows the different queries and $Y$ axis shows the execution time of each query in milliseconds (using log scale). Note that for some queries, the maximum execution time exceeded the time-out threshold which is marked `$0$' in~\reffig{fig::bsbm}. These results suggest that the document-stores with JSON-SNV representation has performance advantages (i.e., around one order of magnitude) over others for subject-subject join query types such as \textbf{Q1}, \textbf{Q2}, and \textbf{Q4} (please see \refsec{subsec::kgqueries} and \reftab{table::queries} for more details about different query types). For example, the query execution time over the BSBM with 10M triples for \textbf{Q1} using JSON-SNV is less than 8 milliseconds using document-stores (i.e., 5.13 and 7.68 milliseconds for MongoDB and Couchbase, respectively) compared with 31.09 (Virtuoso (row)), 27.79 (Virtuoso (column)), 127.87 (Blazegraph), 25.51 (JSON-CNV with MongoDB), 29.91 (JSON-CNV with Couchbase), 2489.9 (JSON-DT with MongoDB), and 3209.45 (JSON-DT with Couchbase) where all execution times are in milliseconds.

\reffig{fig::bsbm} also shows that JSON-CNV outperforms other JSON representations and DMSs by up to one order of magnitude to execute subject-object join queries, namely, \textbf{Q10-11}, \textbf{Q16}, and \textbf{Q18} (please see \reftab{table::queries}). There are no significant differences across JSON representations with respect to \textbf{Q20} as a highly selective query. In our experiments, different DMSs showed performance similarity for executing highly selective queries (aka, point queries). Aside from highly selective queries, JSON-DT performs slower than all other representations and DMSs for almost all other types of queries, especially at scale.

Based on \reffig{fig::bsbm}, the performance advantages of JSON-SNV for executing \textbf{Q1-2} and \textbf{Q4} remained unchanged at different scales. In a similar way, JSON-CNV displays better performance for running \mbox{\textbf{Q10-11}}, \textbf{Q16}, and \textbf{Q18} in both lower and higher scales. The trends related to the performance of different DMSs are almost remained unchanged at different scales.

\begin{figure*}[t]
\begin{adjustbox}{center}
\centering\begin{tabular}{c}
\includegraphics[width=1\textwidth]{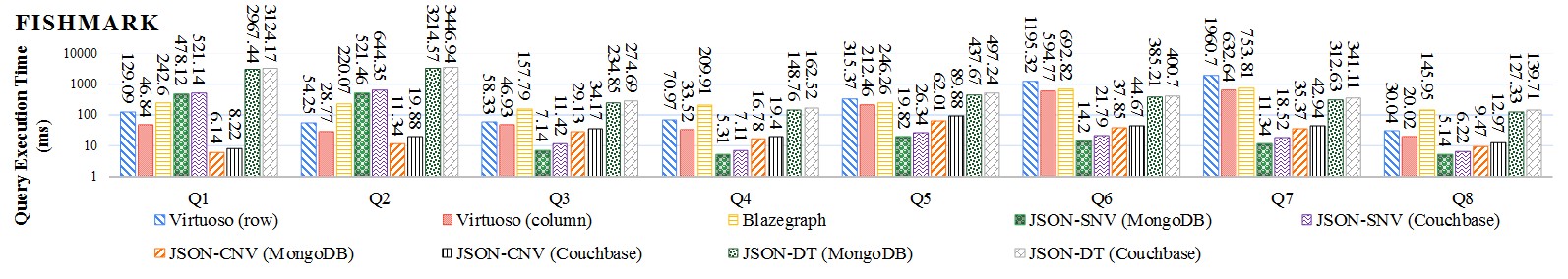}\\
\end{tabular}
\end{adjustbox}
\caption{The FishMark benchmark queries. $X$ axis shows different benchmark queries. $Y$ axis shows the execution time of each query in milliseconds (log scale).}%
\label{fig::fish}%
\end{figure*}

\reffig{fig::watdiv} shows the results obtained from running the WatDiv benchmark queries. They indicate the performance advantages of JSON-SNV for subject-subject join quires (\textbf{Q1-3}) and JSON-CNV for executing subject-object join quires (\textbf{Q4-6}) by around one order of magnitude while more performance similarity can be seen in the execution times of \textbf{Q7-8} as combined queries (see \reftab{table::queries}). \reffig{fig::fish} shows the execution times of FishMark benchmark queries. Document-stores (either MongoDB or Couchbase) outperform others by a factor of 2-8 for executing \textbf{Q1-2} and \textbf{Q4-7} (see \reftab{table::queries}). However, for executing \textbf{Q3} and \textbf{Q8}, the performance differences are not significant across  different representations and DMSs.

\subsection{Analysis}
To better understand the factors contributing to the performance differences, we present our detailed analyses below.

\textbf{JSON-SNV Representation.} In our experiments, the JSON-SNV with document-stores (either MongoDB or Couchbase) outperforms other DMSs by around one order of magnitude to execute subject-subject join queries (please see \refsec{subsec::kgqueries} and \reftab{table::queries}). Virtuoso (both row and column) and Blazegraph typically execute subject-subject join by scanning indexes for each triple pattern separately. The retrieved result of each triple pattern is kept (usually in the main memory) as an intermediary result. These DMSs then join different intermediary results to return the final result. Virtuoso (both row and column) and Blazegraph typically use a hash join algorithm for executing subject-subject joins over the intermediary results. For example, let $L$ and $R$ be two resultsets containing several triples that should be joined based on their subjects. The average time-complexity for typical input data is up to linear complexity $O(|L|+|R|)$ for subject-subject joins in Virtuoso (both row and column) and Blazegraph. However, this complexity is up to $O(\log n)$, where $n$ is the number of unique subjects for document-stores (either MongoDB or Couchbase) using the JSON-SNV representation with a constructed index on subjects. This is because all triples with the same subject have appeared in a single JSON document and the joining of triples with the same subject is equivalent to an index-based look-up querying of a given subject. Therefore, we typically expect to observe better performance from the JSON-SNV data model using document-stores for subject-subject join queries.

\textbf{JSON-CNV Representation.} As presented in (\refsec{sec::results}), the JSON-CNV with document-stores (either MongoDB or Couchbase) outperforms other DMSs by up to one order of magnitude for subject-object join queries (please see \refsec{subsec::kgqueries} and \reftab{table::queries}). Virtuoso (both row and column) and Blazegraph first execute subject-object join by scanning indexes for each triple pattern separately. These DMSs then join different intermediary results (of the index scans) to return the final result. Merge join algorithm is known to be an efficient algorithm to be executed for subject-object join queries~\cite{svenbook}. To the best of our knowledge, Virtuoso (both row and column) and Blazegraph have not implemented the merge join as part of their query processing engines. As a result, these DMSs use Index Nested Loop join to support subject-object queries. The complexity of the Index Nested Loop join is quadratic on the average~\cite{garciabook}. However, the document-stores have paths as first-class citizens of each JSON documents in the JSON-CNV representation (please see \refsec{subsec::CNV}). Therefore, the complexity of executing subject-object join queries using the JSON-CNV with documents-stores is $O(\log n)$ on the average, where $n$ is the number of predicates that are involved in a subject-object join query (i.e., with a constructed index on predicates). This is because all triples which are involved in the path have appeared into a single JSON document and the joining of these triples is equivalent to index-based look-up querying of given predicates. Therefore, we typically expect to observe better performance from the JSON-CNV with document-stores for subject-object join queries.

\textbf{JSON-DT Representation.} This representation is very similar to the natural representation of triples as defined by W3C\footnote{https://www.w3.org/TR/n-triples/}. In general, it does not perform as fast as other representations for most of the queries. This is because the JSON-DT represents KGs using a large number of JSON documents where each document holds a single triple. To execute queries that need filtering of more than a triple, JSON-DT firstly requires group triples together based on their subjects (or predicate/object depending on the query requirement). Such grouping before running each query makes query execution over the JSON-DT slower than the JSON-SNV and the JSON-CNV. However, it performs more similar to others for executing highly selective queries with simpler filtering. Because of the simplicity of the JSON-DT representation, it can be considered as a candidate for data exchange and integration of KG content.

The foregoing explanation suggests that JSON-based representations (with document-stores) perform better than native RDF-stores for subject-subject and subject-object join queries. We took a closer look at other queries like \textbf{WatDiv-Q7-8}. We observed that these queries contain combinations of subject-subject and subject-object joins. In other words, these queries consist of several subject-subject sub-queries connected via some subject-object join queries. The performance of these queries may vary depending on the complexity of such combinations. Therefore, we found that no single DMS displays superior query performance for complex combinations of subject-subject and subject-object queries. Furthermore, we speculate that creating aggressive secondary indexes (aka, compound indexes) over either the JSON-SNV or the JSON-CNV may provide better performance for this type of query as compared to Virtuoso and Blazegraph, especially at scale.


\section{Discussion}
\label{sec::Discussion}

\subsection{Limitations}
Currently, the maximum size of each JSON document in MongoDB is 16MB and in Couchbase is 20MB. These document-stores reject JSON documents when its size exceeds the above values. Technically, the maximum document size in document-stores helps ensure that a single document cannot use an excessive amount of memory. JSON-based representations might be affected negatively by this. In our experiments, there were no cases in which the document size exceeded the maximum value. However, in principle, the size of JSON documents in the JSON-SNV and JSON-CNV may exceed the maximum document size depending on the KG content.

There may be limited redundant or overlapping name/value pairs in the representations. For write-heavy applications, this may cause write latency, however, querying KGs tends to be read-heavy (i.e., scan-mostly). Another issue that remains to be addressed is the automatic conversion of SPARQL to N1QL (i.e., Couchbase's query language) and JavaScript-like (i.e., for MongoDB) queries. In our experiments, we converted the benchmark queries manually and after the execution of each query, we carefully checked to ensure that the output results are correct and exactly the same across different DMSs and representations.

It appears that no single DMS displays superior query performance across the highly selective queries and combinations of subject-subject and subject-object queries. However, the consistency in performance advantages of JSON-SNV representation for subject-subject join queries and JSON-CNV for subject-object join queries was found across different benchmark datasets. These results are likely to be generalizable. However, more experimentation is warranted before we can arrive at any firm conclusions. In our experiments, we had three major query types. There may be other query types that document-stores are unable to execute as efficiently as native RDF-stores.

Document-based DMSs displayed performance advantages for query processing in our experiments. However, in addition to query processing, native RDF-stores (e.g., Virtuoso and Blazegraph) have important functionalities such as inference, incremental truth maintenance, and statement identifiers that document-stores are unable to deliver at this stage.

\subsection{Lessons Learned}

\textbf{Locality.} The main advantage of JSON-based representations for storing KGs is to increase data locality since all the triples related to one resource (i.e., a subject in the JSON-SNV and a path starting from a subject in the JSON-CNV) are located together. We speculate that such locality leads to denser data layout, more CPU cache (i.e., L2 cache) locality and more RAM locality and therefore increased overall performance on typical KG queries especially for subject-subject and subject-object join queries. It appears Virtuoso (especially the row-store version) and Blazegraph do not exploit the data locality and it can incur significant overhead due to buffer pool misses as compared to document stores.

\textbf{Lack of Schema and Heterogeneity of the Underlying Data.} KGs can represent and mix diverse data ranging from \textit{structured} (e.g., DBLP) to \textit{unstructured} (e.g., DBpedia). This \textit{structural flexibility} has contributed to the widespread acceptance of KGs in a variety of domains. Such a structural mixture and flexibility poses a challenge to DMSs for \textit{querying} KG content efficiently since DMSs typically cannot make any \mbox{a priori} assumptions about the structure of the data that is going to be stored. The lack of schema and heterogeneity of the underlying data makes KG querying a harder problem compared to relational data querying. However, we found that the employing of document-stores (e.g., MongoDB or Couchbase) and appropriate JSON-based data models (e.g., the JSON-CNV or JSON-SNV) offer the potential to outperform native RDF-stores (e.g., Blazegraph or Virtuoso) by several orders of magnitude, especially at scale.

\textbf{Intermediary Result.} We note that the performance of KG query evaluation tends to be dominated by the sizes of the query's output and more often its intermediary results. When a query contains more than a triple pattern, DMSs usually have to scan large parts of indexes for each triple pattern and then join the result of these scans. These index scans would produce large intermediary results. We observed that even when the query itself is very selective with small output, the size of the intermediary results can be still very large. The size of the intermediary result challenges native RDF-stores since unlike relational DMSs where most of the joins are foreign-key-based (either one-to-one or one-to-many), in native RDF-stores the index scans can
lead to exponential growth in the size of the intermediate results.

Currently, DMSs usually use either of two techniques data compression or Sideways Information Passing (SIP) to decrease the size of intermediary results. It appears that employing these techniques to decrease intermediary results in native RDF-stores may increase the computation need of query evaluation process for the uncompression or additional filtering (for SIP) requirement. However, our experiments suggest that (especially at scale) the use of JSON-based data models leads to lower memory usage since less memory is allocated to intermediate results.

\textbf{Query Diversity.} It has been noted that beyond the heterogeneity of the underlying data, the KG query workloads also exhibit high variability. Our experiments suggest that the heterogeneity in the KG content query plan can be efficiently addressed by appropriate JSON data models.

\textbf{Textual Values.} We also found that \textit{textual values} tend to be the most prevalent data type in a good proportion of the benchmark KGs. Our results show that the filtering of textual values is time-consuming especially when the text is \textit{lengthy}. Given that typical queries over KGs need to \textit{touch} a large amount of data (e.g., up to 70\% of the whole data), textual value filtering plays a major role in query performance. It may be possible to convert some textual values to numeric values by using hashing and providing dictionaries. Such a conversion usually saves space and allows for more efficient access structures in some cases, e.g., textual resource identifiers and ids. However, this conversion cannot be typically applied on lengthier texts (e.g., comments, reviews, description, etc.) where textual filtering is a \textit{part} of the queries. Fast filtering of textual values poses challenges for efficient KG query execution.

\textbf{Multi-valued Predicates.} We note that Multi-valued predicates tend to appear in almost all subjects in the benchmark KGs, e.g., \reffig{fig::kg} depicts a multi-valued predicate where ``New York'' is an ``instance\_of'': ``city'' and ``metropolis''. We also observed that some predicates are \textit{repeated} for a broad range of subjects, e.g., ``id'', ``name'', ``type'', ``label'', etc. These predicates are less likely to be a part of the filtering in query processing over KGs. However, the final results usually should include them for purposes like applying the sorting modifier over them. Multi-valued or high frequently repeated predicates degrade the effectiveness of index scans through decreasing selectivity of queries. In JSON-based data models, \textit{arrays} are often used for storing multi-valued predicates. Queries for filtering such arrays typically involve long execution time since \textit{sequential scans} of arrays can be time-consuming. However, it appears that filtering multi-valued or high frequently repeated predicates using JSON-based data models with document-stores is still faster than native RDF-stores.

\section{Related Work}
\label{sec::related_work}
Efficient data management plays an important role in unlocking the full potential of Semantic Web applications. Early approaches such as ~\cite{Broek3,Jena17,AntiDan} employed relational database systems to store the Semantic Web datasets. These systems typically store a set of triples by using a relational table with three columns resulting in low implementation overhead. Virtuso~\cite{Virtuso} and RDF3X~\cite{RDF3x} are well-known systems from this category. Abadi et al.~\cite{Dan2007,DanSW} represented some of the first studies in which the importance of data representation for Semantic Web applications using \textit{SQL-based} systems was highlighted and the use of column-oriented DMSs (e.g.,~\cite{CStore}) was proposed. Over time, the emergence (and growing use) of KGs called for systems that can store and evaluate \textit{queries} over them efficiently~\cite{Growing3,Growing2,Growing1,Survey2018}. In response, a variety of DMSs were proposed such as Virtuoso and Blazegraph. As discussed in comprehensive surveys such as~\cite{Kaoudi,Ozsu,Survey2018}, we can classify the previous studies into several categories. We briefly review three major categories, namely, triple-based indexing, infrastructure configuring, and graph processing in the following.

\textbf{Triple-based Indexing.} Virtuoso, HexaStor~\cite{Hexastore}, and the Rya system~\cite{Rya} are three DMSs that are performing mainly based on indexing. For instance, the Rya~\cite{Rya} which is designed on the top of Accumulo~\footnote{https://accumulo.apache.org/} (i.e., a distributed key-value and column-oriented NoSQL store) created indexes on the all permutations of the triple pattern across three separated tables. The permutations include SPO (i.e., S stands for Subjects, P stands for Predicates, and O stands for Objects), POS, and OSP. The effectiveness of triple-based indexing solutions can be limited since querying KGs typically requires touching a large amount of data and complex filtering.

\textbf{Infrastructure Configuring.} JenaHBase~\cite{JenaHbase}, H2RDF~\cite{H2rdf}, and AMADA~\cite{AMADA} are three well-known DMSs that focused mainly on the importance of configurations of underlying infrastructure such as cluster segmentation, communication overhead, and distributed storage layouts. For instance, JenaHBase~\cite{JenaHbase} proposed a custom-built data storage layout for query processing and physical storage.

H2RDF~\cite{H2rdf} combines the HBase\footnote{https://hbase.apache.org/} and the Hadoop\footnote{https://hadoop.apache.org/} framework. H2RDF employed the Hadoop platform to provide a distributed query processing module by launching MapReduce jobs for queries that require touching a large amount of data. H2RDF+~\cite{H2rdfp} extended the H2RDF~\cite{H2rdf} by creating indexes on all permutations of triple patterns in distributed indexing tables. In other words, H2RDF+~\cite{H2rdfp} merged triple-based indexing and infrastructure configuration techniques.

AMADA~\cite{AMADA} also exploited infrastructure configuration techniques by employing cloud computing to store and query data. In particular, AMADA stores the data in the Amazon Simple Storage Service (S3). The S3 interface attaches a URL to each dataset to be used later for the query processing. AMADA used Amazon Simple Queue Service (SQS) and virtual machines within the Amazon Elastic Compute Cloud (EC2) for the query execution.

\textbf{Graph Processing.} Some approaches have applied ideas from the graph processing world to handle KG querying such as Blazegraph, gStore~\cite{gStore}, and \cite{Turbo}. For instance, gStore~\cite{gStore} as a graph-based storage system models KGs as a labeled and directed multi-edge graph. gStore stores the graph by using a disk-based adjacency list table and executes queries by mapping them to a subgraph matching task over the graph. Kim et. al.~\cite{Turbo} considers RDF graphs as labeled graphs and applies subgraph homomorphism methods for query processing. To improve its query performance, it exploits optimization techniques and a Non-Uniform Memory Access (NUMA)-aware parallelism for query processing.

These and related studies like~\cite{SurveySPARQL,ISWC2013} mainly focused on the importance of \textit{configurations} such as the number of storage and computation nodes (i.e., cluster segmentation), storage file systems, and communication efficiency for improving the performance of query execution. However, the \textit{question} of the effects of different data representations on query processing for \textit{Semantic Web applications} has not received much research attention. To the best of our knowledge, our paper is one of the first in which the impacts of JSON data modeling on query processing in the context of KGs and document-stores is investigated.

\section{Conclusions and Future Work}
\label{sec::conclusion}
We have addressed the \textit{question} of the effects of \textit{different JSON-based representations} on query processing for \textit{KGs}. To this end, we have synthesized three distinct JSON-based data modelings for a KG, namely, Subject-based Name/Value (JSON-SNV), Documents of Triples (JSON-DT), and Chain-based Name/Value (JSON-CNV). We analyzed the effects of these distinct representations on KG query performance. Our results showed that JSON-based representations have significant effects on the performance of query evaluation. In particular, we noted that the JSON-SNV with document-stores outperforms others (e.g., Virtuoso and Blazegraph) by around one order of magnitude.

Our results also suggest that queries over KGs are typically memory-bound rather than CPU-bound since there are usually a limited number of arithmetic operations and most of the processing time is spent on random memory accesses. Based on that, we speculate that focusing on effective memory usage, during planning for KGs representations, is often more effective than improving the computation algorithm. We have also discussed why the \textit{JSON-SNV and JSON-CNV} outperformed others.

 Our experience while building the presented JSON-based representations raised several interesting challenges and research directions that need to be addressed in the future, namely:

    \begin{itemize}
        \item Effective representations of multi-valued predicates
        
        \item Serialisations and compression of high-frequency predicates
        
        \item Replication of this research using more datasets and DMSs
        
        \item Decreasing the amount of redundant information in different representations
        
        \item Rewriting SPARQL queries to other declarative query languages such as N1QL automatically
        
    \end{itemize}

\balance
\bibliographystyle{ACM-Reference-Format}
\bibliography{main_bib}


\begin{thebibliography}{60}


\ifx \showCODEN    \undefined \def \showCODEN     #1{\unskip}     \fi
\ifx \showDOI      \undefined \def \showDOI       #1{#1}\fi
\ifx \showISBNx    \undefined \def \showISBNx     #1{\unskip}     \fi
\ifx \showISBNxiii \undefined \def \showISBNxiii  #1{\unskip}     \fi
\ifx \showISSN     \undefined \def \showISSN      #1{\unskip}     \fi
\ifx \showLCCN     \undefined \def \showLCCN      #1{\unskip}     \fi
\ifx \shownote     \undefined \def \shownote      #1{#1}          \fi
\ifx \showarticletitle \undefined \def \showarticletitle #1{#1}   \fi
\ifx \showURL      \undefined \def \showURL       {\relax}        \fi
\providecommand\bibfield[2]{#2}
\providecommand\bibinfo[2]{#2}
\providecommand\natexlab[1]{#1}
\providecommand\showeprint[2][]{arXiv:#2}

\bibitem[\protect\citeauthoryear{Abadi, Marcus, Madden, and Hollenbach}{Abadi
  et~al\mbox{.}}{2007}]%
        {Dan2007}
\bibfield{author}{\bibinfo{person}{Daniel~J. Abadi}, \bibinfo{person}{Adam
  Marcus}, \bibinfo{person}{Samuel~R. Madden}, {and} \bibinfo{person}{Kate
  Hollenbach}.} \bibinfo{year}{2007}\natexlab{}.
\newblock \showarticletitle{Scalable {Semantic Web} Data Management Using
  Vertical Partitioning}. In \bibinfo{booktitle}{\emph{Proc. VLDB Endow.}}
  \bibinfo{pages}{411--422}.
\newblock


\bibitem[\protect\citeauthoryear{Abadi, Marcus, Madden, and Hollenbach}{Abadi
  et~al\mbox{.}}{2009}]%
        {DanSW}
\bibfield{author}{\bibinfo{person}{Daniel~J. Abadi}, \bibinfo{person}{Adam
  Marcus}, \bibinfo{person}{Samuel~R. Madden}, {and} \bibinfo{person}{Kate
  Hollenbach}.} \bibinfo{year}{2009}\natexlab{}.
\newblock \showarticletitle{{SW-Store}: a vertically partitioned {DBMS} for
  {Semantic Web} data management}.
\newblock \bibinfo{journal}{\emph{Proc. VLDB Endow.}} \bibinfo{volume}{18},
  \bibinfo{number}{2} (\bibinfo{year}{2009}), \bibinfo{pages}{385--406}.
\newblock


\bibitem[\protect\citeauthoryear{Abdelaziz, Harbi, Khayyat, and
  Kalnis}{Abdelaziz et~al\mbox{.}}{2017}]%
        {SurveySPARQL}
\bibfield{author}{\bibinfo{person}{Ibrahim Abdelaziz}, \bibinfo{person}{Razen
  Harbi}, \bibinfo{person}{Zuhair Khayyat}, {and} \bibinfo{person}{Panos
  Kalnis}.} \bibinfo{year}{2017}\natexlab{}.
\newblock \showarticletitle{A Survey and Experimental Comparison of Distributed
  {SPARQL} Engines for Very Large {RDF} Data}.
\newblock \bibinfo{journal}{\emph{Proc. VLDB Endow.}} \bibinfo{volume}{10},
  \bibinfo{number}{13} (\bibinfo{year}{2017}), \bibinfo{pages}{2049--2060}.
\newblock


\bibitem[\protect\citeauthoryear{Alexander}{Alexander}{2008}]%
        {AlexRDF}
\bibfield{author}{\bibinfo{person}{Keith Alexander}.}
  \bibinfo{year}{2008}\natexlab{}.
\newblock \showarticletitle{{RDF} in {JSON}: a specification for serialising
  {RDF} in {JSON}}. In \bibinfo{booktitle}{\emph{Proc. of the Int. Workshop on
  Scripting for the {Semantic Web} (SFSW)}}. \bibinfo{pages}{71--77}.
\newblock


\bibitem[\protect\citeauthoryear{Alu{\c{c}}, Hartig, {\"O}zsu, and
  Daudjee}{Alu{\c{c}} et~al\mbox{.}}{2014}]%
        {watdiv}
\bibfield{author}{\bibinfo{person}{G{\"u}ne{\c{s}} Alu{\c{c}}},
  \bibinfo{person}{Olaf Hartig}, \bibinfo{person}{M.~Tamer {\"O}zsu}, {and}
  \bibinfo{person}{Khuzaima Daudjee}.} \bibinfo{year}{2014}\natexlab{}.
\newblock \showarticletitle{Diversified Stress Testing of {RDF} Data Management
  Systems}. In \bibinfo{booktitle}{\emph{Proc. of the Int. {Semantic Web} Conf.
  (ISWC)}}. \bibinfo{pages}{197--212}.
\newblock


\bibitem[\protect\citeauthoryear{Alu\c{c}, \"{O}zsu, and Daudjee}{Alu\c{c}
  et~al\mbox{.}}{2019}]%
        {tamer2019}
\bibfield{author}{\bibinfo{person}{G\"{u}ne\c{s} Alu\c{c}},
  \bibinfo{person}{M.~Tamer \"{O}zsu}, {and} \bibinfo{person}{Khuzaima
  Daudjee}.} \bibinfo{year}{2019}\natexlab{}.
\newblock \showarticletitle{Building Self-clustering RDF Databases Using
  Tunable-LSH}.
\newblock \bibinfo{journal}{\emph{The VLDB Journal}} \bibinfo{volume}{28},
  \bibinfo{number}{2} (\bibinfo{year}{2019}), \bibinfo{pages}{173--195}.
\newblock


\bibitem[\protect\citeauthoryear{Aranda-And\'{u}jar, Bugiotti,
  Camacho-Rodr\'{\i}guez, Colazzo, Goasdou{\'e}, Kaoudi, and
  Manolescu}{Aranda-And\'{u}jar et~al\mbox{.}}{2012}]%
        {AMADA}
\bibfield{author}{\bibinfo{person}{Andr{\'e}s Aranda-And\'{u}jar},
  \bibinfo{person}{Francesca Bugiotti}, \bibinfo{person}{Jes\'{u}s
  Camacho-Rodr\'{\i}guez}, \bibinfo{person}{Dario Colazzo},
  \bibinfo{person}{Fran\c{c}ois Goasdou{\'e}}, \bibinfo{person}{Zoi Kaoudi},
  {and} \bibinfo{person}{Ioana Manolescu}.} \bibinfo{year}{2012}\natexlab{}.
\newblock \showarticletitle{{AMADA}: Web Data Repositories in the {Amazon}
  Cloud}. In \bibinfo{booktitle}{\emph{Proc. of the ACM Int. Conf. on
  Information and Knowledge Management (CIKM)}}. \bibinfo{pages}{2749--2751}.
\newblock


\bibitem[\protect\citeauthoryear{Arenas, Barcel, Libkin, and Murlak}{Arenas
  et~al\mbox{.}}{2014}]%
        {xmlbook}
\bibfield{author}{\bibinfo{person}{Marcelo Arenas}, \bibinfo{person}{Pablo
  Barcel}, \bibinfo{person}{Leonid Libkin}, {and} \bibinfo{person}{Filip
  Murlak}.} \bibinfo{year}{2014}\natexlab{}.
\newblock \bibinfo{booktitle}{\emph{Foundations of Data Exchange}}.
\newblock \bibinfo{publisher}{Cambridge University Press}.
\newblock
\showISBNx{1107016169, 9781107016163}


\bibitem[\protect\citeauthoryear{Atre}{Atre}{2015}]%
        {Medha2}
\bibfield{author}{\bibinfo{person}{Medha Atre}.}
  \bibinfo{year}{2015}\natexlab{}.
\newblock \showarticletitle{Left Bit Right: For {SPARQL} Join Queries with
  {OPTIONAL} Patterns (Left-outer-joins)}. In \bibinfo{booktitle}{\emph{Proc.
  of the ACM Int. Conf. on Management of Data (SIGMOD)}}.
  \bibinfo{pages}{1793--1808}.
\newblock


\bibitem[\protect\citeauthoryear{Bail, Alkiviadous, Parsia, Workman, Harmelen,
  Goncalves, and Garilao}{Bail et~al\mbox{.}}{2012}]%
        {fishmark}
\bibfield{author}{\bibinfo{person}{Samantha Bail}, \bibinfo{person}{Ra
  Alkiviadous}, \bibinfo{person}{Bijan Parsia}, \bibinfo{person}{David
  Workman}, \bibinfo{person}{Van Harmelen}, \bibinfo{person}{Rafael~S.
  Goncalves}, {and} \bibinfo{person}{Cristina Garilao}.}
  \bibinfo{year}{2012}\natexlab{}.
\newblock \showarticletitle{{FishMark}: A linked data application benchmark}.
  In \bibinfo{booktitle}{\emph{Proc. of the Int. Workshop on Scalable and
  High-Performance Semantic Web Systems (SSWS)}}. \bibinfo{pages}{1--15}.
\newblock


\bibitem[\protect\citeauthoryear{Bilidas and Koubarakis}{Bilidas and
  Koubarakis}{2019}]%
        {edbtcore2019}
\bibfield{author}{\bibinfo{person}{Dimitris Bilidas} {and}
  \bibinfo{person}{Manolis Koubarakis}.} \bibinfo{year}{2019}\natexlab{}.
\newblock \showarticletitle{Scalable Parallelization of RDF Joins on Multicore
  Architectures}. In \bibinfo{booktitle}{\emph{Proc. of Int. Conf. on Extending
  Database Technology (EDBT)}}. \bibinfo{pages}{349--360}.
\newblock


\bibitem[\protect\citeauthoryear{Bizer and Schultz}{Bizer and Schultz}{2009}]%
        {bsbm}
\bibfield{author}{\bibinfo{person}{Christian Bizer} {and}
  \bibinfo{person}{Andreas Schultz}.} \bibinfo{year}{2009}\natexlab{}.
\newblock \showarticletitle{The {Berlin SPARQL} Benchmark}.
\newblock \bibinfo{journal}{\emph{Int. J. {Semantic Web} Inf. Syst.}}
  \bibinfo{volume}{5} (\bibinfo{year}{2009}), \bibinfo{pages}{1--24}.
\newblock


\bibitem[\protect\citeauthoryear{Bonifati, Fletcher, Voigt, and
  Yakovets}{Bonifati et~al\mbox{.}}{2018}]%
        {bonifatibook}
\bibfield{author}{\bibinfo{person}{Angela Bonifati}, \bibinfo{person}{George
  H.~L. Fletcher}, \bibinfo{person}{Hannes Voigt}, {and}
  \bibinfo{person}{Nikolay Yakovets}.} \bibinfo{year}{2018}\natexlab{}.
\newblock \bibinfo{booktitle}{\emph{Querying Graphs}}.
\newblock \bibinfo{publisher}{Morgan {\&} Claypool Publishers}.
\newblock


\bibitem[\protect\citeauthoryear{Broekstra, Kampman, and van
  Harmelen}{Broekstra et~al\mbox{.}}{2002}]%
        {Broek3}
\bibfield{author}{\bibinfo{person}{Jeen Broekstra}, \bibinfo{person}{Arjohn
  Kampman}, {and} \bibinfo{person}{Frank van Harmelen}.}
  \bibinfo{year}{2002}\natexlab{}.
\newblock \showarticletitle{Sesame: A Generic Architecture for Storing and
  Querying {RDF} and {RDF} Schema}. In \bibinfo{booktitle}{\emph{Proc. of the
  Int. {Semantic Web} Conf. (ISWC)}}, \bibfield{editor}{\bibinfo{person}{Ian
  Horrocks} {and} \bibinfo{person}{James Hendler}} (Eds.).
  \bibinfo{pages}{54--68}.
\newblock


\bibitem[\protect\citeauthoryear{Chasseur, Li, and Patel}{Chasseur
  et~al\mbox{.}}{2013}]%
        {Jignesh}
\bibfield{author}{\bibinfo{person}{Craig Chasseur}, \bibinfo{person}{Yinan Li},
  {and} \bibinfo{person}{Jignesh~M Patel}.} \bibinfo{year}{2013}\natexlab{}.
\newblock \showarticletitle{Enabling JSON Document Stores in Relational
  Systems.}. In \bibinfo{booktitle}{\emph{WebDB}}, Vol.~\bibinfo{volume}{13}.
  \bibinfo{pages}{14--15}.
\newblock


\bibitem[\protect\citeauthoryear{{Chen}, {Chen}, {Du}, {Zhang}, and
  {Zhou}}{{Chen} et~al\mbox{.}}{2016}]%
        {Growing3}
\bibfield{author}{\bibinfo{person}{J. {Chen}}, \bibinfo{person}{Y. {Chen}},
  \bibinfo{person}{X. {Du}}, \bibinfo{person}{X. {Zhang}}, {and}
  \bibinfo{person}{X. {Zhou}}.} \bibinfo{year}{2016}\natexlab{}.
\newblock \showarticletitle{SEED: A system for entity exploration and debugging
  in large-scale knowledge graphs}. In \bibinfo{booktitle}{\emph{Proc. of the
  IEEE Int. Conf. on Data Engineering (ICDE)}}. \bibinfo{pages}{1350--1353}.
\newblock


\bibitem[\protect\citeauthoryear{Cudr{\'e}-Mauroux, Enchev, Fundatureanu,
  Groth, Haque, Harth, Keppmann, Miranker, Sequeda, and
  Wylot}{Cudr{\'e}-Mauroux et~al\mbox{.}}{2013}]%
        {ISWC2013}
\bibfield{author}{\bibinfo{person}{Philippe Cudr{\'e}-Mauroux},
  \bibinfo{person}{Iliya Enchev}, \bibinfo{person}{Sever Fundatureanu},
  \bibinfo{person}{Paul Groth}, \bibinfo{person}{Albert Haque},
  \bibinfo{person}{Andreas Harth}, \bibinfo{person}{Felix~Leif Keppmann},
  \bibinfo{person}{Daniel Miranker}, \bibinfo{person}{Juan~F. Sequeda}, {and}
  \bibinfo{person}{Marcin Wylot}.} \bibinfo{year}{2013}\natexlab{}.
\newblock \showarticletitle{{NoSQL} Databases for {RDF}: An Empirical
  Evaluation}. In \bibinfo{booktitle}{\emph{Proc. of the Int. {Semantic Web}
  Conf. (ISWC)}}. \bibinfo{pages}{310--325}.
\newblock


\bibitem[\protect\citeauthoryear{Duan, Kementsietsidis, Srinivas, and
  Udrea}{Duan et~al\mbox{.}}{2011}]%
        {IBMapple}
\bibfield{author}{\bibinfo{person}{Songyun Duan}, \bibinfo{person}{Anastasios
  Kementsietsidis}, \bibinfo{person}{Kavitha Srinivas}, {and}
  \bibinfo{person}{Octavian Udrea}.} \bibinfo{year}{2011}\natexlab{}.
\newblock \showarticletitle{Apples and Oranges: A Comparison of {RDF}
  Benchmarks and Real {RDF} Datasets}. In \bibinfo{booktitle}{\emph{Proc. of
  the ACM Int. Conf. on Management of Data (SIGMOD)}}.
  \bibinfo{pages}{145--156}.
\newblock


\bibitem[\protect\citeauthoryear{Erling and Mikhailov}{Erling and
  Mikhailov}{2009}]%
        {Virtuso}
\bibfield{author}{\bibinfo{person}{Orri Erling} {and} \bibinfo{person}{Ivan
  Mikhailov}.} \bibinfo{year}{2009}\natexlab{}.
\newblock \showarticletitle{{RDF} Support in the {Virtuoso DBMS}}. In
  \bibinfo{booktitle}{\emph{Networked Knowledge - Networked Media: Integrating
  Knowledge Management, New Media Technologies and Semantic Systems}}.
  \bibinfo{pages}{7--24}.
\newblock


\bibitem[\protect\citeauthoryear{Fagin, Kolaitis, Miller, and Popa}{Fagin
  et~al\mbox{.}}{2005}]%
        {fagin}
\bibfield{author}{\bibinfo{person}{Ronald Fagin}, \bibinfo{person}{Phokion~G.
  Kolaitis}, \bibinfo{person}{Renée~J. Miller}, {and} \bibinfo{person}{Lucian
  Popa}.} \bibinfo{year}{2005}\natexlab{}.
\newblock \showarticletitle{Data exchange: semantics and query answering}.
\newblock \bibinfo{journal}{\emph{Theoretical Computer Science}}
  \bibinfo{volume}{336}, \bibinfo{number}{1} (\bibinfo{year}{2005}),
  \bibinfo{pages}{89 -- 124}.
\newblock


\bibitem[\protect\citeauthoryear{Fern\'{a}ndez, Mart\'{\i}nez-Prieto, de~la
  Fuente~Redondo, and Guti{\'e}rrez}{Fern\'{a}ndez et~al\mbox{.}}{2018}]%
        {CharacteristicsRDF}
\bibfield{author}{\bibinfo{person}{Javier~D Fern\'{a}ndez},
  \bibinfo{person}{Miguel~A Mart\'{\i}nez-Prieto}, \bibinfo{person}{Pablo de~la
  Fuente~Redondo}, {and} \bibinfo{person}{Claudio Guti{\'e}rrez}.}
  \bibinfo{year}{2018}\natexlab{}.
\newblock \showarticletitle{Characterising {RDF} Data Sets}.
\newblock \bibinfo{journal}{\emph{Journal of Information Science}}
  \bibinfo{volume}{44}, \bibinfo{number}{2} (\bibinfo{year}{2018}),
  \bibinfo{pages}{203--229}.
\newblock


\bibitem[\protect\citeauthoryear{Fernández, Martínez-Prieto, Gutiérrez,
  Polleres, and Arias}{Fernández et~al\mbox{.}}{2013}]%
        {BinaryRDF}
\bibfield{author}{\bibinfo{person}{Javier~D. Fernández},
  \bibinfo{person}{Miguel~A. Martínez-Prieto}, \bibinfo{person}{Claudio
  Gutiérrez}, \bibinfo{person}{Axel Polleres}, {and} \bibinfo{person}{Mario
  Arias}.} \bibinfo{year}{2013}\natexlab{}.
\newblock \showarticletitle{Binary {RDF} representation for publication and
  exchange {(HDT)}}.
\newblock \bibinfo{journal}{\emph{Journal of Web Semantics}}
  \bibinfo{volume}{19} (\bibinfo{year}{2013}), \bibinfo{pages}{22 -- 41}.
\newblock


\bibitem[\protect\citeauthoryear{Fionda and Pirr\`{o}}{Fionda and
  Pirr\`{o}}{2017}]%
        {QR}
\bibfield{author}{\bibinfo{person}{Valeria Fionda} {and}
  \bibinfo{person}{Giuseppe Pirr\`{o}}.} \bibinfo{year}{2017}\natexlab{}.
\newblock \showarticletitle{Explaining and Querying Knowledge Graphs by
  Relatedness}.
\newblock \bibinfo{journal}{\emph{Proc. VLDB Endow.}} \bibinfo{volume}{10},
  \bibinfo{number}{12} (\bibinfo{year}{2017}), \bibinfo{pages}{1913--1916}.
\newblock


\bibitem[\protect\citeauthoryear{Fuxman and Miller}{Fuxman and Miller}{2009}]%
        {Fuxman}
\bibfield{author}{\bibinfo{person}{Ariel Fuxman} {and}
  \bibinfo{person}{Ren{\'e}e~J. Miller}.} \bibinfo{year}{2009}\natexlab{}.
\newblock \bibinfo{booktitle}{\emph{Schema Mapping}}.
\newblock \bibinfo{publisher}{Springer US}, \bibinfo{pages}{2481--2488}.
\newblock


\bibitem[\protect\citeauthoryear{Gad-Elrab, Stepanova, Urbani, and
  Weikum}{Gad-Elrab et~al\mbox{.}}{2019}]%
        {Exfact}
\bibfield{author}{\bibinfo{person}{Mohamed~H. Gad-Elrab},
  \bibinfo{person}{Daria Stepanova}, \bibinfo{person}{Jacopo Urbani}, {and}
  \bibinfo{person}{Gerhard Weikum}.} \bibinfo{year}{2019}\natexlab{}.
\newblock \showarticletitle{{ExFaKT}: A Framework for Explaining Facts over
  Knowledge Graphs and Text}. In \bibinfo{booktitle}{\emph{Proc. of the ACM
  Int. Conf. on Web Search and Data Mining (WSDM)}}. \bibinfo{pages}{87--95}.
\newblock


\bibitem[\protect\citeauthoryear{Garcia-Molina}{Garcia-Molina}{2008}]%
        {garciabook}
\bibfield{author}{\bibinfo{person}{Hector Garcia-Molina}.}
  \bibinfo{year}{2008}\natexlab{}.
\newblock \bibinfo{booktitle}{\emph{Database systems: the complete book}}.
\newblock \bibinfo{publisher}{Pearson Education India}.
\newblock


\bibitem[\protect\citeauthoryear{Groppe}{Groppe}{2011}]%
        {svenbook}
\bibfield{author}{\bibinfo{person}{Sven Groppe}.}
  \bibinfo{year}{2011}\natexlab{}.
\newblock \bibinfo{booktitle}{\emph{Data management and query processing in
  {Semantic Web} databases}}.
\newblock \bibinfo{publisher}{Springer Science \& Business Media}.
\newblock


\bibitem[\protect\citeauthoryear{Gu, Zhou, Cheng, Li, Pan, and Qu}{Gu
  et~al\mbox{.}}{2019}]%
        {SemanticNet}
\bibfield{author}{\bibinfo{person}{Yu Gu}, \bibinfo{person}{Tianshuo Zhou},
  \bibinfo{person}{Gong Cheng}, \bibinfo{person}{Ziyang Li},
  \bibinfo{person}{Jeff~Z. Pan}, {and} \bibinfo{person}{Yuzhong Qu}.}
  \bibinfo{year}{2019}\natexlab{}.
\newblock \showarticletitle{Relevance Search over Schema-Rich Knowledge
  Graphs}. In \bibinfo{booktitle}{\emph{Proc. of the ACM Int. Conf. on Web
  Search and Data Mining (WSDM)}}. \bibinfo{pages}{114--122}.
\newblock


\bibitem[\protect\citeauthoryear{Gutierrez, Hurtado, and Vaisman}{Gutierrez
  et~al\mbox{.}}{2011}]%
        {Formaldef2}
\bibfield{author}{\bibinfo{person}{Claudio Gutierrez}, \bibinfo{person}{Carlos
  Hurtado}, {and} \bibinfo{person}{Alejandro Vaisman}.}
  \bibinfo{year}{2011}\natexlab{}.
\newblock \showarticletitle{{RDFS} Update: From Theory to Practice}. In
  \bibinfo{booktitle}{\emph{The {Semanic Web}: Research and Applications}}.
  \bibinfo{pages}{93--107}.
\newblock


\bibitem[\protect\citeauthoryear{{Jayaram}, {Khan}, {Li}, {Yan}, and
  {Elmasri}}{{Jayaram} et~al\mbox{.}}{2015}]%
        {Growing2}
\bibfield{author}{\bibinfo{person}{N. {Jayaram}}, \bibinfo{person}{A. {Khan}},
  \bibinfo{person}{C. {Li}}, \bibinfo{person}{X. {Yan}}, {and}
  \bibinfo{person}{R. {Elmasri}}.} \bibinfo{year}{2015}\natexlab{}.
\newblock \showarticletitle{Querying Knowledge Graphs by Example Entity
  Tuples}.
\newblock \bibinfo{journal}{\emph{IEEE Trans. on Knowledge and Data Engineering
  (TKDE)}} \bibinfo{volume}{27}, \bibinfo{number}{10} (\bibinfo{year}{2015}),
  \bibinfo{pages}{2797--2811}.
\newblock


\bibitem[\protect\citeauthoryear{Kaoudi and Manolescu}{Kaoudi and
  Manolescu}{2015}]%
        {Kaoudi}
\bibfield{author}{\bibinfo{person}{Zoi Kaoudi} {and} \bibinfo{person}{Ioana
  Manolescu}.} \bibinfo{year}{2015}\natexlab{}.
\newblock \showarticletitle{{RDF} in the clouds: a survey}.
\newblock \bibinfo{journal}{\emph{Proc. VLDB Endow.}} \bibinfo{volume}{24},
  \bibinfo{number}{1} (\bibinfo{year}{2015}), \bibinfo{pages}{67--91}.
\newblock


\bibitem[\protect\citeauthoryear{Khadilkar, Kantarcioglu, Thuraisingham, and
  Castagna}{Khadilkar et~al\mbox{.}}{2012}]%
        {JenaHbase}
\bibfield{author}{\bibinfo{person}{Vaibhav Khadilkar}, \bibinfo{person}{Murat
  Kantarcioglu}, \bibinfo{person}{Bhavani Thuraisingham}, {and}
  \bibinfo{person}{Paolo Castagna}.} \bibinfo{year}{2012}\natexlab{}.
\newblock \showarticletitle{Jena-HBase: A Distributed, Scalable and Efficient
  {RDF} Triple Store}. In \bibinfo{booktitle}{\emph{Proc. of the Int. {Semantic
  Web} Conf. (ISWC)}}. \bibinfo{pages}{85--88}.
\newblock


\bibitem[\protect\citeauthoryear{Kim, Shin, Han, Hong, and Chafi}{Kim
  et~al\mbox{.}}{2015}]%
        {Turbo}
\bibfield{author}{\bibinfo{person}{Jinha Kim}, \bibinfo{person}{Hyungyu Shin},
  \bibinfo{person}{Wook-Shin Han}, \bibinfo{person}{Sungpack Hong}, {and}
  \bibinfo{person}{Hassan Chafi}.} \bibinfo{year}{2015}\natexlab{}.
\newblock \showarticletitle{Taming Subgraph Isomorphism for {RDF} Query
  Processing}.
\newblock \bibinfo{journal}{\emph{Proc. VLDB Endow.}} \bibinfo{volume}{8},
  \bibinfo{number}{11} (\bibinfo{year}{2015}), \bibinfo{pages}{1238--1249}.
\newblock


\bibitem[\protect\citeauthoryear{Lockard, Dong, Einolghozati, and
  Shiralkar}{Lockard et~al\mbox{.}}{2018}]%
        {Arash}
\bibfield{author}{\bibinfo{person}{Colin Lockard}, \bibinfo{person}{Xin~Luna
  Dong}, \bibinfo{person}{Arash Einolghozati}, {and} \bibinfo{person}{Prashant
  Shiralkar}.} \bibinfo{year}{2018}\natexlab{}.
\newblock \showarticletitle{CERES: Distantly Supervised Relation Extraction
  from the Semi-structured Web}.
\newblock \bibinfo{journal}{\emph{Proc. VLDB Endow.}} \bibinfo{volume}{11},
  \bibinfo{number}{10} (\bibinfo{year}{2018}), \bibinfo{pages}{1084--1096}.
\newblock


\bibitem[\protect\citeauthoryear{{McBride}}{{McBride}}{2002}]%
        {Jena17}
\bibfield{author}{\bibinfo{person}{B. {McBride}}.}
  \bibinfo{year}{2002}\natexlab{}.
\newblock \showarticletitle{Jena: a {Semantic Web} toolkit}.
\newblock \bibinfo{journal}{\emph{IEEE Internet Computing}}
  \bibinfo{volume}{6}, \bibinfo{number}{6} (\bibinfo{year}{2002}),
  \bibinfo{pages}{55--59}.
\newblock


\bibitem[\protect\citeauthoryear{Melo and Paulheim}{Melo and Paulheim}{2017}]%
        {Growing1}
\bibfield{author}{\bibinfo{person}{Andr{\'e} Melo} {and} \bibinfo{person}{Heiko
  Paulheim}.} \bibinfo{year}{2017}\natexlab{}.
\newblock \showarticletitle{Synthesizing Knowledge Graphs for Link and Type
  Prediction Benchmarking}. In \bibinfo{booktitle}{\emph{Proc. of the Int.
  European Semantic Web Conf. (ESWC)}}. \bibinfo{pages}{136--151}.
\newblock


\bibitem[\protect\citeauthoryear{Mohanty, Ramanath, Yahya, and Weikum}{Mohanty
  et~al\mbox{.}}{2019}]%
        {edbtkg2019}
\bibfield{author}{\bibinfo{person}{Madhulika Mohanty}, \bibinfo{person}{Maya
  Ramanath}, \bibinfo{person}{Mohamed Yahya}, {and} \bibinfo{person}{Gerhard
  Weikum}.} \bibinfo{year}{2019}\natexlab{}.
\newblock \showarticletitle{{Spec-QP}: Speculative Query Planning for Joins
  over Knowledge Graphs}. In \bibinfo{booktitle}{\emph{Proc. of Int. Conf. on
  Extending Database Technology (EDBT)}}. \bibinfo{pages}{61--72}.
\newblock


\bibitem[\protect\citeauthoryear{Neumann and Weikum}{Neumann and
  Weikum}{2009}]%
        {Thomasindex}
\bibfield{author}{\bibinfo{person}{Thomas Neumann} {and}
  \bibinfo{person}{Gerhard Weikum}.} \bibinfo{year}{2009}\natexlab{}.
\newblock \showarticletitle{Scalable Join Processing on Very Large RDF Graphs}.
  In \bibinfo{booktitle}{\emph{Proc. of the ACM Int. Conf. on Management of
  Data (SIGMOD)}}. \bibinfo{pages}{627--640}.
\newblock


\bibitem[\protect\citeauthoryear{Neumann and Weikum}{Neumann and
  Weikum}{2010}]%
        {RDF3x}
\bibfield{author}{\bibinfo{person}{Thomas Neumann} {and}
  \bibinfo{person}{Gerhard Weikum}.} \bibinfo{year}{2010}\natexlab{}.
\newblock \showarticletitle{The {RDF-3X} engine for scalable management of
  {RDF} data}.
\newblock \bibinfo{journal}{\emph{Proc. VLDB Endow.}} \bibinfo{volume}{19},
  \bibinfo{number}{1} (\bibinfo{year}{2010}), \bibinfo{pages}{91--113}.
\newblock


\bibitem[\protect\citeauthoryear{Oulabi and Bizer}{Oulabi and Bizer}{2019}]%
        {edbtyago2019}
\bibfield{author}{\bibinfo{person}{Yaser Oulabi} {and}
  \bibinfo{person}{Christian Bizer}.} \bibinfo{year}{2019}\natexlab{}.
\newblock \showarticletitle{Extending Cross-Domain Knowledge Bases with Long
  Tail Entities using Web Table Data}. In \bibinfo{booktitle}{\emph{Proc. of
  Int. Conf. on Extending Database Technology (EDBT)}}.
  \bibinfo{pages}{385--396}.
\newblock


\bibitem[\protect\citeauthoryear{\"{O}zsu}{\"{O}zsu}{2016}]%
        {Ozsu}
\bibfield{author}{\bibinfo{person}{M.~Tamer \"{O}zsu}.}
  \bibinfo{year}{2016}\natexlab{}.
\newblock \showarticletitle{A Survey of {RDF} Data Management Systems}.
\newblock \bibinfo{journal}{\emph{Frontiers of Computer Science}}
  \bibinfo{volume}{10}, \bibinfo{number}{3} (\bibinfo{year}{2016}),
  \bibinfo{pages}{418--432}.
\newblock


\bibitem[\protect\citeauthoryear{Papailiou, Konstantinou, Tsoumakos, Karras,
  and Koziris}{Papailiou et~al\mbox{.}}{2013}]%
        {H2rdfp}
\bibfield{author}{\bibinfo{person}{Nikolaos Papailiou},
  \bibinfo{person}{Ioannis Konstantinou}, \bibinfo{person}{Dimitrios
  Tsoumakos}, \bibinfo{person}{Panagiotis Karras}, {and}
  \bibinfo{person}{Nectarios Koziris}.} \bibinfo{year}{2013}\natexlab{}.
\newblock \showarticletitle{{H2RDF+}: High-performance distributed joins over
  large-scale {RDF} graphs}.
\newblock \bibinfo{journal}{\emph{Proc. of the IEEE Int. Conf. on Big Data}}
  (\bibinfo{year}{2013}), \bibinfo{pages}{255--263}.
\newblock


\bibitem[\protect\citeauthoryear{Papailiou, Konstantinou, Tsoumakos, and
  Koziris}{Papailiou et~al\mbox{.}}{2012}]%
        {H2rdf}
\bibfield{author}{\bibinfo{person}{Nikolaos Papailiou},
  \bibinfo{person}{Ioannis Konstantinou}, \bibinfo{person}{Dimitrios
  Tsoumakos}, {and} \bibinfo{person}{Nectarios Koziris}.}
  \bibinfo{year}{2012}\natexlab{}.
\newblock \showarticletitle{{H2RDF}: Adaptive Query Processing on {RDF} Data in
  the Cloud.}. In \bibinfo{booktitle}{\emph{Proc. of the Int. Conf. on World
  Wide Web (WWW)}}. \bibinfo{pages}{397--400}.
\newblock


\bibitem[\protect\citeauthoryear{Peng, Zou, {\"{O}}zsu, Chen, and Zhao}{Peng
  et~al\mbox{.}}{2016}]%
        {Peng}
\bibfield{author}{\bibinfo{person}{Peng Peng}, \bibinfo{person}{Lei Zou},
  \bibinfo{person}{M.~Tamer {\"{O}}zsu}, \bibinfo{person}{Lei Chen}, {and}
  \bibinfo{person}{Dongyan Zhao}.} \bibinfo{year}{2016}\natexlab{}.
\newblock \showarticletitle{Processing {SPARQL} queries over distributed {RDF}
  graphs}.
\newblock \bibinfo{journal}{\emph{Proc. VLDB Endow.}} \bibinfo{volume}{25},
  \bibinfo{number}{2} (\bibinfo{year}{2016}), \bibinfo{pages}{243--268}.
\newblock


\bibitem[\protect\citeauthoryear{P{\'e}rez, Arenas, and Gutierrez}{P{\'e}rez
  et~al\mbox{.}}{2006}]%
        {Formaldef1}
\bibfield{author}{\bibinfo{person}{Jorge P{\'e}rez}, \bibinfo{person}{Marcelo
  Arenas}, {and} \bibinfo{person}{Claudio Gutierrez}.}
  \bibinfo{year}{2006}\natexlab{}.
\newblock \showarticletitle{Semantics and Complexity of {SPARQL}}. In
  \bibinfo{booktitle}{\emph{Proc. of the Int. Semantic Web Conf. (ISWC)}}.
  \bibinfo{pages}{30--43}.
\newblock


\bibitem[\protect\citeauthoryear{Punnoose, Crainiceanu, and Rapp}{Punnoose
  et~al\mbox{.}}{2015}]%
        {Rya}
\bibfield{author}{\bibinfo{person}{Roshan Punnoose}, \bibinfo{person}{Adina
  Crainiceanu}, {and} \bibinfo{person}{David Rapp}.}
  \bibinfo{year}{2015}\natexlab{}.
\newblock \showarticletitle{{SPARQL} in the Cloud Using {Rya}}.
\newblock \bibinfo{journal}{\emph{Inf. Syst.}}  \bibinfo{volume}{48}
  (\bibinfo{year}{2015}), \bibinfo{pages}{181--195}.
\newblock


\bibitem[\protect\citeauthoryear{Saeedi, Peukert, and Rahm}{Saeedi
  et~al\mbox{.}}{2018}]%
        {Erham1}
\bibfield{author}{\bibinfo{person}{Alieh Saeedi}, \bibinfo{person}{Eric
  Peukert}, {and} \bibinfo{person}{Erhard Rahm}.}
  \bibinfo{year}{2018}\natexlab{}.
\newblock \showarticletitle{Using Link Features for Entity Clustering in
  Knowledge Graphs}. In \bibinfo{booktitle}{\emph{The Semantic Web}}.
  \bibinfo{pages}{576--592}.
\newblock


\bibitem[\protect\citeauthoryear{Sakr, Wylot, Mutharaju, Le~Phuoc, and
  Fundulaki}{Sakr et~al\mbox{.}}{2018}]%
        {sakrbook}
\bibfield{author}{\bibinfo{person}{Sherif Sakr}, \bibinfo{person}{Marcin
  Wylot}, \bibinfo{person}{Raghava Mutharaju}, \bibinfo{person}{Danh Le~Phuoc},
  {and} \bibinfo{person}{Irini Fundulaki}.} \bibinfo{year}{2018}\natexlab{}.
\newblock \bibinfo{booktitle}{\emph{Linked Data: Storing, Querying, and
  Reasoning}}.
\newblock \bibinfo{publisher}{Springer}.
\newblock


\bibitem[\protect\citeauthoryear{Saleem, Sz\'{a}rnyas, Conrads, Bukhari,
  Mehmood, and Ngonga~Ngomo}{Saleem et~al\mbox{.}}{2019}]%
        {saleem}
\bibfield{author}{\bibinfo{person}{Muhammad Saleem}, \bibinfo{person}{G\'{a}bor
  Sz\'{a}rnyas}, \bibinfo{person}{Felix Conrads}, \bibinfo{person}{Syed
  Ahmad~Chan Bukhari}, \bibinfo{person}{Qaiser Mehmood}, {and}
  \bibinfo{person}{Axel-Cyrille Ngonga~Ngomo}.}
  \bibinfo{year}{2019}\natexlab{}.
\newblock \showarticletitle{How Representative Is a SPARQL Benchmark? An
  Analysis of RDF Triplestore Benchmarks}. In \bibinfo{booktitle}{\emph{Proc.
  of the Int. Conf. on World Wide Web (WWW)}}. \bibinfo{pages}{1623–1633}.
\newblock


\bibitem[\protect\citeauthoryear{Schenkel, Hose, and Harth}{Schenkel
  et~al\mbox{.}}{2014}]%
        {lodbook}
\bibfield{author}{\bibinfo{person}{Ralf Schenkel}, \bibinfo{person}{Katja
  Hose}, {and} \bibinfo{person}{Andreas Harth}.}
  \bibinfo{year}{2014}\natexlab{}.
\newblock \bibinfo{booktitle}{\emph{Linked Data Management}}.
\newblock \bibinfo{publisher}{Taylor \& Francis}.
\newblock


\bibitem[\protect\citeauthoryear{Sidirourgos, Goncalves, Kersten, Nes, and
  Manegold}{Sidirourgos et~al\mbox{.}}{2008}]%
        {AntiDan}
\bibfield{author}{\bibinfo{person}{Lefteris Sidirourgos},
  \bibinfo{person}{Romulo Goncalves}, \bibinfo{person}{Martin Kersten},
  \bibinfo{person}{Niels Nes}, {and} \bibinfo{person}{Stefan Manegold}.}
  \bibinfo{year}{2008}\natexlab{}.
\newblock \showarticletitle{Column-store Support for {RDF} Data Management: Not
  All Swans Are White}.
\newblock \bibinfo{journal}{\emph{Proc. VLDB Endow.}} \bibinfo{volume}{1},
  \bibinfo{number}{2} (\bibinfo{year}{2008}), \bibinfo{pages}{1553--1563}.
\newblock


\bibitem[\protect\citeauthoryear{Stonebraker, Abadi, Batkin, Chen, Cherniack,
  Ferreira, Lau, Lin, Madden, O'Neil, O'Neil, Rasin, Tran, and
  Zdonik}{Stonebraker et~al\mbox{.}}{2005}]%
        {CStore}
\bibfield{author}{\bibinfo{person}{Mike Stonebraker},
  \bibinfo{person}{Daniel~J. Abadi}, \bibinfo{person}{Adam Batkin},
  \bibinfo{person}{Xuedong Chen}, \bibinfo{person}{Mitch Cherniack},
  \bibinfo{person}{Miguel Ferreira}, \bibinfo{person}{Edmond Lau},
  \bibinfo{person}{Amerson Lin}, \bibinfo{person}{Sam Madden},
  \bibinfo{person}{Elizabeth O'Neil}, \bibinfo{person}{Pat O'Neil},
  \bibinfo{person}{Alex Rasin}, \bibinfo{person}{Nga Tran}, {and}
  \bibinfo{person}{Stan Zdonik}.} \bibinfo{year}{2005}\natexlab{}.
\newblock \showarticletitle{{C-store}: A Column-oriented {DBMS}}. In
  \bibinfo{booktitle}{\emph{Proc. VLDB Endow.}} \bibinfo{pages}{553--564}.
\newblock


\bibitem[\protect\citeauthoryear{Subercaze, Gravier, Chevalier, and
  Laforest}{Subercaze et~al\mbox{.}}{2016}]%
        {Inferray}
\bibfield{author}{\bibinfo{person}{Julien Subercaze},
  \bibinfo{person}{Christophe Gravier}, \bibinfo{person}{Jules Chevalier},
  {and} \bibinfo{person}{Frederique Laforest}.}
  \bibinfo{year}{2016}\natexlab{}.
\newblock \showarticletitle{{Inferray}: Fast In-memory {RDF} Inference}.
\newblock \bibinfo{journal}{\emph{Proc. VLDB Endow.}} \bibinfo{volume}{9},
  \bibinfo{number}{6} (\bibinfo{year}{2016}), \bibinfo{pages}{468--479}.
\newblock


\bibitem[\protect\citeauthoryear{Wang, Zhang, Shi, Jiao, Hassanzadeh, Zou, and
  Wangz}{Wang et~al\mbox{.}}{2015}]%
        {JSONTree}
\bibfield{author}{\bibinfo{person}{Lanjun Wang}, \bibinfo{person}{Shuo Zhang},
  \bibinfo{person}{Juwei Shi}, \bibinfo{person}{Limei Jiao},
  \bibinfo{person}{Oktie Hassanzadeh}, \bibinfo{person}{Jia Zou}, {and}
  \bibinfo{person}{Chen Wangz}.} \bibinfo{year}{2015}\natexlab{}.
\newblock \showarticletitle{Schema Management for Document Stores}.
\newblock \bibinfo{journal}{\emph{Proc. VLDB Endow.}} \bibinfo{volume}{8},
  \bibinfo{number}{9} (\bibinfo{year}{2015}), \bibinfo{pages}{922--933}.
\newblock


\bibitem[\protect\citeauthoryear{Weiss, Karras, and Bernstein}{Weiss
  et~al\mbox{.}}{2008}]%
        {Hexastore}
\bibfield{author}{\bibinfo{person}{Cathrin Weiss}, \bibinfo{person}{Panagiotis
  Karras}, {and} \bibinfo{person}{Abraham Bernstein}.}
  \bibinfo{year}{2008}\natexlab{}.
\newblock \showarticletitle{Hexastore: Sextuple Indexing for {Semantic Web}
  Data Management}.
\newblock \bibinfo{journal}{\emph{Proc. VLDB Endow.}} \bibinfo{volume}{1},
  \bibinfo{number}{1} (\bibinfo{year}{2008}), \bibinfo{pages}{1008--1019}.
\newblock


\bibitem[\protect\citeauthoryear{Wu, Fujiwara, Yamamoto, Bolleman, and
  Yamaguchi}{Wu et~al\mbox{.}}{2014}]%
        {biobench}
\bibfield{author}{\bibinfo{person}{Hongyan Wu}, \bibinfo{person}{Toyofumi
  Fujiwara}, \bibinfo{person}{Yasunori Yamamoto}, \bibinfo{person}{Jerven
  Bolleman}, {and} \bibinfo{person}{Atsuko Yamaguchi}.}
  \bibinfo{year}{2014}\natexlab{}.
\newblock \showarticletitle{{BioBenchmark Toyama} 2012: an evaluation of the
  performance of triple stores on biological data}.
\newblock \bibinfo{journal}{\emph{Journal of Biomedical Semantics}}
  \bibinfo{volume}{5}, \bibinfo{number}{1} (\bibinfo{year}{2014}),
  \bibinfo{pages}{32--43}.
\newblock


\bibitem[\protect\citeauthoryear{Wylot, Hauswirth, Cudr{\'e}-Mauroux, and
  Sakr}{Wylot et~al\mbox{.}}{2018}]%
        {Survey2018}
\bibfield{author}{\bibinfo{person}{Marcin Wylot}, \bibinfo{person}{Manfred
  Hauswirth}, \bibinfo{person}{Philippe Cudr{\'e}-Mauroux}, {and}
  \bibinfo{person}{Sherif Sakr}.} \bibinfo{year}{2018}\natexlab{}.
\newblock \showarticletitle{{RDF} Data Storage and Query Processing Schemes: A
  Survey}.
\newblock \bibinfo{journal}{\emph{ACM Comput. Surv.}} \bibinfo{volume}{51},
  \bibinfo{number}{4} (\bibinfo{year}{2018}), \bibinfo{pages}{84:1--84:36}.
\newblock


\bibitem[\protect\citeauthoryear{Zhang, Paudel, Zhang, Bernstein, and
  Chen}{Zhang et~al\mbox{.}}{2019}]%
        {GenKG}
\bibfield{author}{\bibinfo{person}{Wen Zhang}, \bibinfo{person}{Bibek Paudel},
  \bibinfo{person}{Wei Zhang}, \bibinfo{person}{Abraham Bernstein}, {and}
  \bibinfo{person}{Huajun Chen}.} \bibinfo{year}{2019}\natexlab{}.
\newblock \showarticletitle{Interaction Embeddings for Prediction and
  Explanation in Knowledge Graphs}. In \bibinfo{booktitle}{\emph{Proc. of the
  ACM Int. Conf. on Web Search and Data Mining (WSDM)}}.
  \bibinfo{pages}{96--104}.
\newblock


\bibitem[\protect\citeauthoryear{Zheng, Yu, Zou, and Cheng}{Zheng
  et~al\mbox{.}}{2018}]%
        {QA}
\bibfield{author}{\bibinfo{person}{Weiguo Zheng}, \bibinfo{person}{Jeffrey~Xu
  Yu}, \bibinfo{person}{Lei Zou}, {and} \bibinfo{person}{Hong Cheng}.}
  \bibinfo{year}{2018}\natexlab{}.
\newblock \showarticletitle{Question Answering over Knowledge Graphs: Question
  Understanding via Template Decomposition}.
\newblock \bibinfo{journal}{\emph{Proc. VLDB Endow.}} \bibinfo{volume}{11},
  \bibinfo{number}{11} (\bibinfo{year}{2018}), \bibinfo{pages}{1373--1386}.
\newblock


\bibitem[\protect\citeauthoryear{Zou, \"{O}zsu, Chen, Shen, Huang, and
  Zhao}{Zou et~al\mbox{.}}{2014}]%
        {gStore}
\bibfield{author}{\bibinfo{person}{Lei Zou}, \bibinfo{person}{M.~Tamer
  \"{O}zsu}, \bibinfo{person}{Lei Chen}, \bibinfo{person}{Xuchuan Shen},
  \bibinfo{person}{Ruizhe Huang}, {and} \bibinfo{person}{Dongyan Zhao}.}
  \bibinfo{year}{2014}\natexlab{}.
\newblock \showarticletitle{gStore: A Graph-based {SPARQL} Query Engine}.
\newblock \bibinfo{journal}{\emph{The VLDB Journal}} \bibinfo{volume}{23},
  \bibinfo{number}{4} (\bibinfo{year}{2014}), \bibinfo{pages}{565--590}.
\newblock


\end{thebibliography}

\end{document}